\documentclass{article}

\usepackage{arxiv}

\usepackage[utf8]{inputenc} 
\usepackage[T1]{fontenc}    
\usepackage{hyperref}       
\usepackage{url}            
\usepackage{booktabs}       
\usepackage{amsfonts}       
\usepackage{nicefrac}       
\usepackage{microtype}      
\usepackage{lipsum}		
\usepackage{graphicx}
\usepackage{natbib}
\usepackage{doi}
\usepackage{bm}
\usepackage{amsmath}
\usepackage{float}
\usepackage{multirow}
\usepackage{graphicx}
\usepackage{nicematrix} 
\usepackage{mathabx}

\usepackage{xcolor}
\usepackage{soul} 

\definecolor{eGreen}{rgb}{.057, .549,.065}
\definecolor{burntorange}{rgb}{0.8, 0.33, 0.0}

\newtheorem{remark}{Remark}[section]

\DeclareMathOperator*{\argmax}{arg\,max}

\title{Bayesian Estimation of Covariate Assisted Principal Regression for brain functional connectivity}

\date{} 					

\author{Hyung Park}

\author{ \href{https://orcid.org/0000-0002-8994-9583}{\includegraphics[scale=0.06]{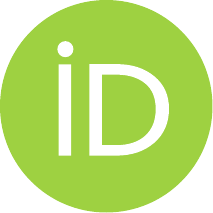}\hspace{1mm}Hyung G. Park } 
\\
	Division of Biostatistics, 
	Department of Population Health \\
	New York University  School of Medicine\\
	New York, NY 10016 \\
	\texttt{parkh15@nyu.edu} \\
}

\renewcommand{\headeright}{Accepted Version}
\renewcommand{\undertitle}{Manuscript}
\renewcommand{\shorttitle}{\textit{arXiv} Bayesian Covariate Assisted Principal Regression }

\hypersetup{
pdftitle={A template for the arxiv style},
pdfsubject={q-bio.NC, q-bio.QM},
pdfauthor={David S.~Hippocampus, Elias D.~Striatum},
pdfkeywords={Bayesian single index models, Heterogeneous treatment effects, Precision medicine},
}

\begin{document}
\maketitle

\begin{abstract} 


This paper presents a Bayesian reformulation of covariate-assisted principal (CAP) regression of \cite{Zhao2021}, which aims to identify components in the covariance of response signal 
that are associated with covariates in a regression framework. We introduce a geometric formulation and reparameterization of individual covariance matrices in their tangent space. By mapping the covariance matrices to the tangent space, we leverage Euclidean geometry to perform posterior inference. This approach enables joint estimation of all parameters and uncertainty quantification within a unified framework, 
fusing dimension reduction for covariance matrices with regression model estimation. We validate the proposed method through simulation studies and apply it to analyze associations between covariates and brain functional connectivity, utilizing data from the Human Connectome Project.

\end{abstract}


 \section{Introduction}
 This paper reformulates covariate-assisted principal (CAP) regression of \cite{Zhao2021} in the Bayesian paradigm. 
The approach identifies covariate-relevant components of the covariance of multivariate response data. 
Specifically,  the method estimates a set of linear projections of multivariate response signals, whose variance is  related to external covariates. 
In neuroscience, there is interest in analyzing 
statistical dependency between time-series of brain signals from distinct regions of the brain, 
which we refer to as functional connectivity (FC) 
\citep[][]{Fox2015,Lindquist2008,Monti2014, Fornito2013, Fornito2012}. 
The brain signals underling FC are multivariate, and 
each brain activity is considered relative to others \citep{Varoquaux2010} in analyzing FC, as this statistical dependency  
is related with behavioral characteristics (covariates). 
This paper develops a Bayesian approach to conducting supervised dimension reduction for the response signals, 
to analyze the association between external covariates and the FC characterized by the multivariate signals'  covariances. 

Typically, 
the first step to analyze brain  FC is to define a set of nodes corresponding to spatial regions of interest (ROIs), 
where each node is associated with its own time course of imaging data. 
Then the network connections (or an ``edge'' structure between the nodes) 
  are subsequently 
  estimated  
based on the statistical dependency between each of the nodes' time course \citep{Friston2011, vanderHeuvel2010}. 
FC networks have been inferred using Pearson's correlation coefficients \citep{Hutchison2013} 
and also with partial correlations in the context of Gaussian graphical models \citep{Whittaker1990, Hinne2014} summarized in the precision or inverse covariance matrix. 
In recent years, there has been a focus on subject-level graphical models where the node-to-node dependencies vary with respect to subject-level covariates. 
This line of research involves methods to estimate or test group-specific graphs \citep{Guo2011, Danaher2014, Xia2017, Xia2018, Narayan2015, Durante2018, Xia2015, Cai2016, Saegusa2016, Peterson2015, Tan2017, Lin2017}   
as well as general Gaussian graphical models for graph edges that allow both continuous and discrete covariates, estimated based on trees  \citep{Liu2010}, kernels \citep{Kolar2010, Lee2018}, 
linear or additive regression \citep{Ni2019, Wang2022, Zhang2022}. 
However, like other standard node-wise regression methods \citep[e.g.,][]{Leday2017, Meinshausen2006, Peng2009, Kolar2010, Cheng2014, Ha2021} 
in Gaussian graphical models, 
these approaches focus on edge detection (i.e., estimation of the off-diagonal elements) rather than estimating the full precision or covariance matrix  and do not explicitly constrain positive definiteness of precision or covariance matrices. 
Works on general tensor outcome regression \citep{Sun2017,Li2017,Lock2018} also do not generally guarantee the positive definiteness of the outcomes.  
While the problem of dimension reduction of individual covariances has been studied in brain dynamic connectivity analysis \citep{Dai2020}, problems in 
computer vision    \citep[][]{Harandi2017, Li2018, Gao2023} and brain computer interfaces \citep[][]{Xie2017, Davoudi2017}  as well as multi-group covariance estimation \citep{Flury1984,Flury1986b,Boik2002, Pourahmadi2007, Hoff2009, Franks2019}, 
covariate information was not utilized in conducting dimension reduction, or it views the data at the group level, 
which does not account for subject-level heterogeneity in the brain networks. 
 Gaussian graphical models have been applied to study brain connectivity networks in fMRI data \citep[e.g.,][]{Li2018Solea, Zhang2020}, however, the focus was on analyzing connectivity networks, without explicitly considering their relationship with subject-level covariates.

In this paper, in line with the covariance regression literatures 
 \citep[see, e.g.,][]{Hoff2012, Fox2015, Zou2017, Pourahmadi2011, Varoquaux2010, Zhao2021, Zhao2021b, Zhao2022, Engle1995, Fong2006}, 
we will frame the problem of analyzing FC  as modeling of heteroscedasticity, 
 i.e., estimating a covariance function $\bm{\Sigma}_{\bm{x}}  = \mbox{var}[\bm{Y}|\bm{x}]$  
across a range of values for an explanatory $\bm{x}$-variable. In contrast to the approach developed in \cite{Zhao2021} where each projection vector for  $\bm{\Sigma}_i$ 
 is estimated sequentially and in \cite{Franks2022} where statistical inference is conducted conditionally on the estimated dimension-reduced subspace, the proposed framework allows coherent and simultaneous inference on all model parameters within the Bayesian paradigm. 
 
 One typical approach to 
associating  brain FC with behavior is to take a 
 massive univariate test approach that relates each connectivity matrix element with subject-level covariates \citep[e.g.,][]{Grillon2013, Woodward2011}. 
However, this ``massive edgewise regression''   lacks statistical power, as it (i)  ignores dependencies among the connectivity elements; 
and (ii)  involves  quadratically increasing number of regressions that exacerbate the problem of multiple testing.
On the other hand, multivariate methods such as principal component analysis (PCA) 
as considered in \cite{Crainiceanu2011} 
consider the data from all ROIs at once, reducing the dimensionality of the original outcome to a smaller number of ``networks'' components, 
 however, these common components may be associated with small eigenvalues, or the corresponding eigenvalues may not be associated with covariates.

 The outcome data of interest are multivariate 
time-series resting-state fMRI (rs-fMRI) 
 data in $\mathbb{R}^p$ measured simultaneously across the $p$ ROIs (or parcels) 
 defined based on an anatomical parcellation \citep{Eickhoff2018} or  ``network nodes''  \citep{Smith2012}  derived from a data-driven algorithm such as 
independent component analysis (ICA) \citep{Calhoun2009, Smith2013}.  
As in \cite{Seiler2017}, 
we will apply the Bayesian CAP regression to data from the Human Connectome Project (HCP) \citep{VanEssen2013} to compare  short sleepers (i.e., $\le 6$ hours) with conventional sleepers  (i.e., $7$ to $9$ hours)  with respect to their  FC.

\section{Method} 
 
\subsection{Covariance regression models} 
     
 We consider $n$ subjects, with subject-specific covariances for brain activity time series from $p$ ROIs   
$\{ \bm{\Sigma}_i \in \mathbb{R}^{p \times p}, i=1,\ldots,n \}$. 
The space of valid covariance matrices $\bm{\Sigma}_i \in \mathbb{R}^{p \times p}$ 
is the space  of symmetric positive definite (SPD) matrices, denoted as $\mbox{Sym}_p^{+}$ in this paper. The rs-fMRI time-series for a given subject $i$ are drawn from a Gaussian distribution: 
$\bm{Y}_{it}  \sim N(\bm{\mu}_i, \bm{\Sigma}_i)$ with 
$\bm{\mu}_i \in \mathbb{R}^p$ and  
$\bm{\Sigma}_i \in \mbox{Sym}_p^{+}$. 
For centered data, the mean  $\bm{\mu}_i = \bm{0}$, 
and the covariance $\bm{\Sigma}_i $ captures FC. 
 Without loss of generality,  we assume  that the observed signal is mean-centered so that $\sum_{t=1}^{T_i} \bm{Y}_{it} = \bm{0} \in \mathbb{R}^p$ for each subject $(i=1,\ldots, n)$, as our focus is on FC characterized by the covariance between the brain signals. 
We observed $\bm{Y}_{it}$ over $T_i$ time points for each subject $i$ $(i=1,\ldots,n)$ along with subject-level vectors of covariates $\bm{x}_i \in \mathbb{R}^q$ $(i=1,\ldots,n)$. 

In this paper, 
instead of directly modeling the subject-specific covariances $\bm{\Sigma}_i = \mbox{cov}(\bm{Y}_{it})$ (as in \cite{Seiler2017,  Fox2015, Zou2017})   
in which most of the covariance heterogeneity may be unrelated with $\bm{x}_i$, 
 we aim to extract a lower dimensional component whose  covariance heterogeneity 
is related with $\bm{x}_i$. We will characterize this lower dimensional structure by a dimension reducing matrix $\bm{\Gamma} \in \mathbb{R}^{p \times d}$ where $\bm{\Gamma}^\top \bm{\Gamma} = \bm{I}_{d}$ (i.e., $\bm{\Gamma}$ is in a Stiefel manifold) with $d \le p$. Specifically, we consider a latent factor model for $\bm{Y}_{it}$
\begin{equation} \label{lfm}
\begin{aligned}
\bm{Y}_{it} 
 = \bm{\Gamma} \bm{\Psi}_i^{\frac{1}{2}}  \bm{s}_{it}  + \bm{L}_i \bm{\epsilon}_{it}    
\end{aligned}
\end{equation}
with latent factors  $\bm{s}_{it}   \sim N(\bm{0}, \bm{I}_{d})$ and 
$\bm{\epsilon}_{it}  \sim N(\bm{0}, \bm{I}_{p-d})$,   
of dimensions $d$ and $p-d$, respectively, where 
\begin{equation} \label{lfm2}
\bm{\Psi}_i^{\frac{1}{2}}  =  \exp\left(\mbox{diag}((\bm{B} \bm{x}_i + \bm{z}_i )/2)\right)
\end{equation} 
models 
  the $\bm{x}$-related heteroscedasticity along the projection directions  $ \bm{\Gamma} \in \mathbb{R}^{p \times d}$. 
  In (\ref{lfm2}),  
 $\mbox{diag}((\bm{B} \bm{x}_i + \bm{z}_i )/2) \in \mathbb{R}^{d \times d}$ is a 
  diagonal matrix, where its diagonal elements are given by 
 a linear predictor vector $(\bm{B} \bm{x}_i + \bm{z}_i )/2 \in \mathbb{R}^d$. 
In (\ref{lfm}), 
$ \bm{\Gamma} \in \mathbb{R}^{p \times d}$ specifies the Principal Directions of Covariance (PDCs) of 
$\bm{Y}_{it} $ 
related with $\bm{x}_i$, 
whereas  
the other 
orthogonal components $\bm{L}_i \in \mathbb{R}^{p \times (p-d)}$, which satisfy $\bm{L}_i \perp \bm{\Gamma}$,   
are included to account for the ``noise''   directions and magnitudes of the heteroscedasticity that are unrelated with $\bm{x}_i$.

In (\ref{lfm2}),  the matrix 
$\bm{B} = [\bm{\beta}_0, \check{\bm{B}} ] \in \mathbb{R}^{d \times q}$ 
(where $\bm{\beta}_0 \in \mathbb{R}^d$ represents the intercept) is a regression coefficient matrix 
that relates  
$\bm{x}_i \in \mathbb{R}^q$ 
(with its first element being 1) 
  to the subject-level outcome covariance $\bm{\Sigma}_i$. Under model (\ref{lfm}),   
the subject-level covariance is given by  
\begin{equation} \label{cov.model}
\bm{\Sigma}_i   =
  \bm{\Gamma} \bm{\Psi}_i
   \bm{\Gamma}^\top 
 +  \bm{L}_i  \bm{L}_i^\top, 
 \end{equation}
  that decomposes the individual covariance matrices  $\bm{\Sigma}_i$ into two components, covariates related 
  and unrelated,  
 a principal factor decomposition of  $\bm{\Sigma}_i$. 
In (\ref{cov.model}), unlike the more general structure on 
 $\bm{L}_i \bm{L}_i^\top$ 
 whose variability is unrelated with $\bm{x}_i$, 
 the  PDCs $\bm{\Gamma}$ serve as 
features (i.e., ``subnetworks'') that we expect 
 to be consistent across subjects. 
Along $\bm{\Gamma}$,
 model (\ref{lfm2}) incorporates subject-level random effects 
 $\bm{z}_i \sim N(\bm{0}, \bm{\Omega})$ to capture additional heteroscedasticity not captured by $\bm{x}_i$.    
 In model (\ref{lfm2}), 
the diagonality of the $d \times d$ 
core tensor $\bm{\Psi}_i \in \mbox{Sym}_d^{+}$ 
  is needed as an identifiability condition, 
  since any non-diagonal SPD $\tilde{\bm{\Psi}}_i$ can be diagonalized by its normalized eigenvectors
  $\tilde{\bm{A}} \in \mathbb{R}^{d \times d}$ (assuming common eigenvectors $\tilde{\bm{A}}$ for $\tilde{\bm{\Psi}}_i$ across subjects), 
and $\bm{\Gamma} \tilde{\bm{A}} \in \mathbb{R}^{p \times d}$ 
   can instead be used  as the orthonormal dimension reduction matrix.  
   While we impose 
  the diagonality of $\bm{\Psi}_i$, 
   we allow $\bm{z}_i \sim N(\bm{0}, \bm{\Omega})$, where $\bm{\Omega}$    may have off-diagonal elements 
   that allow residual correlation 
    in the the projected signals 
    $\bm{\Gamma}^\top \bm{Y}_{it} \in \mathbb{R}^d$  
    beyond what is modeled by common covariates
   $\bm{x}_i$.

 \begin{remark}
 The covariance model (\ref{lfm}) and (\ref{lfm2}) should be distinguished from the principal component (PC) regression that relates $\bm{x}_i$ with the PCs $\bm{\Gamma}^\top \bm{Y}_{it} \in \mathbb{R}^d$, 
 as our interest is in studying the association between the covariate $\bm{x}_i$ with the variance of the components (i.e., heteroscedasticity), 
  rather than with 
 the components $\bm{\Gamma}^\top \bm{Y}_{it} \in \mathbb{R}^d$ themselves.   
\end{remark}

For a multivariate outcome signal $\bm{Y}_{it} \in \mathbb{R}^p$ at time point $t$ for subject $i$,   
\cite{Seiler2017} utilized 
a heteroscedesticity model, 
$\mbox{cov}(\bm{Y}_{it}) = \bm{B}\bm{x}_i \bm{x}_i^\top \bm{B}^\top + \sigma^2 \bm{I}_{p}$ $(t=1,\ldots,T_i) (i=1,\ldots,n)$, 
where the outcome 
 covariance matrix $ \bm{\Sigma}_i = \mbox{cov}(\bm{Y}_{it})$ is modeled by a quadratic function of 
$ \bm{B}\bm{x}_i \in \mathbb{R}^p$, where $\bm{B} \in \mathbb{R}^{p \times q}$ is 
 the regression coefficient associated with $\bm{x}_i \in \mathbb{R}^q$, 
 and $ \sigma^2 >0$. 
However, this model is quite restrictive, as
its outer product term 
$\bm{B}\bm{x}_i \bm{x}_i^\top \bm{B}^\top \in \mathbb{R}^{p \times p}$  is of rank 1, 
and the noise covariance term $\sigma^2 \bm{I}_{p}$ is diagonal with independent variances. 
On the other hand, 
model (\ref{cov.model}) 
identifies a covariate associated rank-$d$ (where $d \ge 1)$ structure via $\bm{\Gamma}$ and allows a less 
restrictive noise covariance structure, 
which makes the covariance modeling 
 with $\bm{x}_i$ more flexible than that of \cite{Seiler2017}. In particular, the outcome dimension reduction via $\bm{\Gamma}$ implicit in model (\ref{cov.model})  
offers computational advantages   
through working with low dimensional ($d$-by-$d$) covariances (rather than full $p$-by-$p$ covariances), 
that can be particularly advantageous when the number of within-subject time points ($T_i$) is relatively small compared to the signal dimension $p$.  
The general outer product approach proposed by \cite{Hoff2012} replaces 
$\sigma^2 \bm{I}_{p}$ by a $p \times p$ SPD matrix, requiring a large number of parameters 
 (that can scale quadratically in $p$). The approaches proposed in \cite{Fox2015, Zou2017} also similarly model the whole $p \times p$ matrix  
$\bm{\Sigma}_i$, which may make the interpretation challenging for large matrices \citep{Zhao2021}.  

\cite{Zhao2021} considered CAP regression, 
 $\mbox{var}(\bm{\gamma}^{(k)\top} \bm{Y}_{it}) = \exp( \bm{x}_i^\top \bm{\beta}^{(k)}),$  
where the PDCs 
$\bm{\gamma}^{(k)} \in \mathbb{R}^p$ $(k=1,\ldots,d)$ 
are sequentially estimated  
subject to identifiability constraints 
$\bm{\gamma}^{(k)\top}\widebar{\bm{\Sigma}} \bm{\gamma}^{(k)} =1$ 
(in which $\widebar{\bm{\Sigma}}$ is a  $p \times p$ covariance representative of the overall study population) 
and 
$\bm{\gamma}^{(k)} \perp \bm{\gamma}^{(k^\prime)}$  $(k \ne k^\prime)$. 
However, under a sequential optimization framework, joint inference on the outcome projection matrix 
$\bm{\Gamma} = [\bm{\gamma}^{(1)}, \ldots, \bm{\gamma}^{(d)}] \in \mathbb{R}^{p \times d}$ 
and the regression coefficient $\bm{B} = [\bm{\beta}^{(1)}, \ldots, \bm{\beta}^{(d)}]  \in \mathbb{R}^{q \times d}$ is not straightforward, 
and 
thus, 
  \cite{Zhao2021,Zhao2021b, Zhao2022} 
  conducted 
bootstrap-based statistical inference only on the coefficients $\bm{B}$, and not on $\bm{\Gamma}$. 
On  the other hand,
the proposed model (\ref{lfm}), coupled with the core tensor model (\ref{lfm2}), further 
accounts for the additional heteroscedasticity in the projected outcomes by using subject-level random effets $\bm{z}_i$ 
to relax the model assumption, 
while simultaneously modeling all the relevant  parameters $(\bm{\Gamma}, \bm{B}, \bm{\Psi}_i, \bm{\Omega})$, 
allowing for more coherent downstream analysis that improves the model interpretability 
which we will discuss in Section 4.  
 
\subsection{Tangent space parametrization of dimension-reduced covariance} 
  
  Due to the constraint 
 $\bm{v}^\top\bm{\Sigma}_i \bm{v} \ge 0$ for all nonzero $\bm{v} \in \mathbb{R}^p$, 
   the space $\mbox{Sym}_p^{+}$ of covariance matrices $\{\bm{\Sigma}_i\}$ 
 forms a curved manifold which does not conform to Euclidean geometry;  for example, 
the negative of a SPD matrix and some linear combinations of SPD matrices are not SPD \citep{Schwartzman2016}. 
Thus, analyzing $\bm{\Sigma}_i$ in the Euclidean vector space is not adequate 
to capture the curved nature of PDCs, 
and leads to a biased estimation of PDCs  \citep{Zhao2021}. 
However, $\mbox{Sym}_p^{+}$ is a Riemannian manifold  under the affine-invariant Riemannian metric (AIRM) 
  \citep{Pennec2006},
   whose 
 tangent space forms a vector space. 
We will use a Riemannian parametrization of SPD matrices in estimating the PDCs in this paper. 
A tangent space projection requires selection of a reference point 
 that is close to $\bm{\Sigma}_i$ $(i=1,\ldots,n)$ to be projected. 
A sensible reference point on $\mbox{Sym}_p^{+}$ is a mean of $\bm{\Sigma}_i$ $(i=1,\ldots,n)$, 
 denoted as $\widebar{\bm{\Sigma}} \in \mbox{Sym}_p^{+}$. 
We will use the matrix whitening transport of \cite{Ng2016}  
to bring the covariances $\bm{\Sigma}_i$ $(i=1,\ldots,n)$ close to $\bm{I}_{p}$,  
by applying matrix whitening based on $\widebar{\bm{\Sigma}}$.  
The resulting whitened covariances 
$\widebar{\bm{\Sigma}}^{-\frac{1}{2}} \bm{\Sigma}_i \widebar{\bm{\Sigma}}^{-\frac{1}{2}}$ 
would be close to the identity matrix $\bm{I}_{p}$, 
at which we can construct a common tangent  space for projection.  
    
     \begin{remark}
Here we briefly review 
  some relevant concepts of Riemannian geometry.  
   Let $\bm{A} \in \mbox{Sym}_p^{+}$, and  
$T_{\bm{A}}(\mbox{Sym}_p^{+})$ be the tangent space at $\bm{A}$. 
  Given two tangent vectors 
    $\bm{X}_1, \bm{X}_2 \in T_{\bm{A}}(\mbox{Sym}_p^{+})$ 
    at $\bm{A}$,   the AIRM inner product is  
    $\langle \bm{X}_1, \bm{X}_2 \rangle_{\bm{A}} = 
    \mbox{tr}(\bm{A}^{-1} \bm{X}_1 \bm{A}^{-1} \bm{X}_2)$.  
Given $\bm{X} \in T_{\bm{A}}(\mbox{Sym}_p^{+})$, 
there is a unique geodesic denoted as 
$\gamma(t) \in \mbox{Sym}_p^{+}$ such that 
$\gamma(0)  = \bm{A}$ and 
$\gamma^\prime(0)  = \bm{X}$, 
\begin{equation} \label{geodesic}
\gamma(t) = 
    \mbox{Exp}_{\bm{A}}(t \bm{X}) 
    =
   \bm{A}^{\frac{1}{2}} \exp(t \bm{A}^{-\frac{1}{2}} \bm{X} \bm{A}^{-\frac{1}{2}} )    \bm{A}^{\frac{1}{2}}   
\end{equation} 
that connects $\bm{A}$ to a point $\bm{B} = \gamma(1) \in \mbox{Sym}_p^{+}$ 
 when evaluated at $t=1$. 
For $\bm{X} \in T_{\bm{A}}(\mbox{Sym}_p^{+})$, 
 the Exponential map, 
 defined as 
 $\mbox{Exp}_{\bm{A}}(\bm{X}) := \gamma(1)  \in \mbox{Sym}_p^{+}$,  
 projects the given $\bm{X}$ 
to a point $\bm{B}  \in \mbox{Sym}_p^{+}$, 
in such a way  that 
 the $\bm{A}$ and $\bm{X}$ distance on the tangent plane 
 is the same as that between $\bm{A}$ 
and $\bm{B}$ on the manifold. 
 The (AIRM) Log map,
which is the inverse mapping of  $\mbox{Exp}_{\bm{A}}(\bm{X})$, 
   projects the point $\bm{B} \in \mbox{Sym}_p^{+}$ 
 back to the tangent  vector,   
\begin{equation} \label{LogMap}
\bm{X} \ =  \mbox{Log}_{\bm{A}}(\bm{B}) 
  =
  \bm{A}^{\frac{1}{2}} \log(\bm{A}^{-\frac{1}{2}} \bm{B} \bm{A}^{-\frac{1}{2}} )    \bm{A}^{\frac{1}{2}} \  \in \  T_{\bm{A}}(\mbox{Sym}_p^{+}), 
 \end{equation}
and we can re-express the geodesic (\ref{geodesic}) as 
$\gamma(t) = 
    \mbox{Exp}_{\bm{A}}(t \ \mbox{Log}_{\bm{A}}(\bm{B})),  t \in [0, 1]$. 
  The corresponding geodesic distance between $\bm{A}$ and $\bm{B}$ is  
  $d(\bm{A}, \bm{B}) = \langle \mbox{Log}_{\bm{A}}(\bm{B}), \mbox{Log}_{\bm{A}}(\bm{B}) \rangle_{\bm{A}}^{\frac{1}{2}} 
  = 
  \lVert 
  \log(\bm{A}^{-\frac{1}{2}}\bm{B}\bm{A}^{-\frac{1}{2}})
   \rVert_F, 
  $ 
  where $\lVert 
\cdot
   \rVert_F$ is 
  the Frobenius norm. 
     \end{remark}

  In this paper, 
 for each dimension reducing matrix $\bm{\Gamma} \in \mathbb{R}^{p \times d}$, 
  we will use $\widebar{\bm{\Psi}} := \bm{\Gamma}^\top \widebar{\bm{\Sigma}} \bm{\Gamma} \in \mbox{Sym}_d^{+}$, 
  where 
  $\widebar{\bm{\Sigma}}$ is a fixed representative  population level covariance, 
  to ``whiten''  the individual level dimension-reduced covariances $\bm{\Psi}_{i}= \bm{\Gamma}^\top \bm{\Sigma}_i \bm{\Gamma} \in \mbox{Sym}_d^{+}$ $(i=1,\ldots,n)$ of model 
(\ref{cov.model}). 
Specifically, 
we will normalize $\bm{\Psi}_i$ by 
$\widebar{\bm{\Psi}}^{-\frac{1}{2}}$ (where $\widebar{\bm{\Psi}}^{-\frac{1}{2}}$ is computed based on the eigendecomposition of $\widebar{\bm{\Psi}} = \bm{\Gamma}^\top \widebar{\bm{\Sigma}} \bm{\Gamma}$),  
  so that the resulting individiual ``whitened'' SPD 
   $\bm{\Psi}_{i}^\ast := 
    \widebar{\bm{\Psi}}^{-\frac{1}{2}} \bm{\Psi}_{i} \widebar{\bm{\Psi}}^{-\frac{1}{2}} 
    = \bm{\Gamma}^\top \widebar{\bm{\Sigma}}^{-\frac{1}{2}} \bm{\Sigma}_i  \widebar{\bm{\Sigma}}^{-\frac{1}{2}} \bm{\Gamma}$   $(i=1,\ldots,n)$ 
is close to the identity matrix $\bm{I}_{d}$.   
We will parametrize these $\bm{\Psi}_{i}^\ast$  
$(i=1,\ldots,n)$ 
 in the tangent space at $\bm{I}_{d}$, by projecting 
 $ \bm{\Psi}_{i}^\ast$ at $ \bm{I}_{d}$ using the Log map, 
\begin{equation} \label{tsp}
 \mbox{Log}_{\bm{I}_{d}}(\bm{\Psi}_{i}^\ast)
 \ = \ 
\log(\bm{\Psi}_{i}^\ast) = 
   \log(\widebar{\bm{\Psi}}^{-\frac{1}{2}} \bm{\Psi}_{i} \widebar{\bm{\Psi}}^{-\frac{1}{2}}) 
  \quad (=  \phi_{\widebar{\bm{\Psi}}}(\bm{\Psi}_{i})), 
 \end{equation}
 locally mapping  the bipoint $\widebar{\bm{\Psi}}, \bm{\Psi}_{i}  \in \mbox{Sym}_d^{+} \times \mbox{Sym}_d^{+}$ 
 to an element in the tangent space at $\bm{I}_{d}$. 
 For notational convenience, in (\ref{tsp})  
 let us denote the Log map, 
  $\log(\widebar{\bm{\Psi}}^{-\frac{1}{2}} \bm{\Psi}_{i} \widebar{\bm{\Psi}}^{-\frac{1}{2}})$      
  given $\widebar{\bm{\Psi}}$, 
as 
  $\phi_{\widebar{\bm{\Psi}}}(\bm{\Psi}_{i}) \in \mathbb{R}^{d \times d}$, 
which is no longer linked by the positive definiteness constraint \citep{Pervaiz2020} and forms a vector space.  
Then, treating $\bm{\Psi}_{i}$ as a local perturbation of $\widebar{\bm{\Psi}}$ in tangent space, 
we model $\phi_{\widebar{\bm{\Psi}}}(\bm{\Psi}_{i})$ in (\ref{tsp}) by a linear model of the form, 
\begin{equation} \label{lp}
  \phi_{\widebar{\bm{\Psi}}}(\bm{\Psi}_{i}) = \mbox{diag}(\tilde{\bm{B}}\bm{x}_i + \tilde{\bm{z}}_i)
 \end{equation} 
 where the linear predictor 
 $\tilde{\bm{B}}\bm{x}_i + \tilde{\bm{z}}_i \in \mathbb{R}^d$ lies in (unrestricted) Euclidean vector space. 
 Upon parametrizing $\phi_{\widebar{\bm{\Psi}}}(\bm{\Psi}_{i})$ (with appropriate priors on 
 $\tilde{\bm{B}}$ and $\tilde{\bm{z}}_i \sim N(\bm{0}, \bm{\Omega})$), 
we will re-map 
 these covariate-parametrized objects $\phi_{\widebar{\bm{\Psi}}}(\bm{\Psi}_{i})$ in (\ref{lp}) to the original space in  $\mbox{Sym}_d^{+}$,  
 by first taking Exponential map, $\mbox{Exp}(\phi_{\widebar{\bm{\Psi}}}(\bm{\Psi}_{i})) =\exp(\phi_{\widebar{\bm{\Psi}}}(\bm{\Psi}_{i}))$ (i.e., taking  (\ref{geodesic}) at 
 $t=1$ and 
  $\bm{A} = \bm{I}_{d}$)   
 and then translating it back to the base point $ \bm{\Gamma}^\top \widebar{\bm{\Sigma}}   \bm{\Gamma}$ 
through  ``de-whitening''  with  $\widebar{\bm{\Psi}} = \bm{\Gamma}^\top \widebar{\bm{\Sigma}}   \bm{\Gamma}$, 
yielding 
\begin{equation}\label{parametrization}
  \bm{\Psi}_{i} = \exp(\phi_{\widebar{\bm{\Psi}}}(\bm{\Psi}_{i}) )  \bm{\Gamma}^\top \widebar{\bm{\Sigma}}   \bm{\Gamma}
 \end{equation} 
 which 
completes our parameterization of the core tensor $\bm{\Psi}_{i}$ in  (\ref{cov.model}).  
To define the mapping (\ref{tsp}), 
 we select $\widebar{\bm{\Sigma}}$ to represent  an estimate of the Euclidean average of $\bm{\Sigma}_i$. Among 
examined estimators 
 in previous works \citep{Pervaiz2020,Dadi2019} 
this choice of $\widebar{\bm{\Sigma}}$ 
showed stable performance across various scenarios. 
We set $\widebar{\bm{\Sigma}} = 
\frac{1}{n} \sum_{i=1}^{n} \widehat{\bm{\Sigma}}_i$, 
where
$\widehat{\bm{\Sigma}}_i = \frac{1}{T_i} \sum_{l=1}^{T_i} \bm{Y}_{it} \bm{Y}_{it}^\top$.

\subsection{Posterior inference} \label{posterior.inference}

  \subsubsection{Prior and likelihood specification} 
  
 We perform posterior inference on the tangent space parameterized model  (\ref{lp}),
  which will be mapped to parametrization (\ref{lfm2}).  
 Let $\mathcal{D}$ 
  represent the observed data   and 
   $\bm{\Psi}$ 
   denote the collection $\{ \bm{\Psi}_i \}_{i=1}^n$,  
   and  let $\bm{Y}_i = \{\bm{Y}_{i1},\ldots, \bm{Y}_{iT_i}\}$. 
  The posterior of 
   parameters 
   $(\bm{\Gamma}, \bm{\Psi}, \tilde{\bm{B}}, \bm{\Omega})$  
    can be expressed as the  
   the product of 
 a prior and the likelihood, 
\begin{equation} \label{posterior} 
p(\bm{\Gamma}, \bm{\Psi}, \tilde{\bm{B}}, \bm{\Omega} | \mathcal{D}) 
 \quad \propto \quad 
p(\bm{\Gamma}, \bm{\Psi}, \tilde{\bm{B}}, \bm{\Omega}) \
\prod_{i=1}^n p\left(\bm{Y}_{i} |  \bm{\Gamma}, \bm{\Psi}, \tilde{\bm{B}}, \bm{\Omega} \right). 
\end{equation} 
The covariate relevant component likelihood 
for subject $i$ under (\ref{lfm}) is 
\begin{equation} \label{likelihood}
\begin{aligned}
p(\bm{Y}_{i} | \bm{\Gamma},  \bm{\Psi}, \tilde{\bm{B}}, \bm{\Omega}) \ 
&\propto \
| \bm{\Gamma} \bm{\Psi}_{i} \bm{\Gamma}^\top +  \bm{L}_i \bm{L}_i^\top|^{-T_i/2} 
\exp(-\frac{1}{2} \sum_{t=1}^{T_i} \bm{Y}_{it}^\top ( \bm{\Gamma} \bm{\Psi}_{i}^{-1} \bm{\Gamma}^\top +  \bm{L}_i (\bm{L}_i^\top \bm{L}_i)^{-1} \bm{L}_i^\top) \bm{Y}_{it} )  \\
&\propto \ 
| \bm{\Psi}_{i} |^{-T_i/2} 
\exp(-\frac{1}{2} \sum_{t=1}^{T_i} \bm{Y}_{it}^\top  \bm{\Gamma} \bm{\Psi}_{i}^{-1} \bm{\Gamma}^\top \bm{Y}_{it}) \
| \bm{\Xi}_{i} |^{-T_i/2} 
\exp(-\frac{1}{2} \sum_{t=1}^{T_i} \bm{Y}_{it}^\top \bm{L}_i \bm{\Xi}_{i}^{-1} \bm{L}_i^\top \bm{Y}_{it}) 
\\
&\propto \ 
| \bm{\Psi}_{i} |^{-T_i/2} 
\exp(-\frac{1}{2} \sum_{t=1}^{T_i} \mbox{tr}(\bm{Y}_{it}^\top  \bm{\Gamma} \bm{\Psi}_{i}^{-1} \bm{\Gamma}^\top \bm{Y}_{it})) \\
&\propto \ 
| \exp(\phi_{\widebar{\bm{\Psi}}}(\bm{\Psi}_{i}) ) 
|^{-T_i/2}
\exp(- \frac{1}{2}
\sum_{t=1}^{T_i} 
  \mbox{tr}
( \bm{Y}_{it}    \widebar{\bm{\Sigma}}^{-\frac{1}{2}}  
\bm{\Gamma} 
\left(\exp(\phi_{\widebar{\bm{\Psi}}}(\bm{\Psi}_{i}) )\right)^{-1} 
    \bm{\Gamma}^\top 
 \widebar{\bm{\Sigma}}^{-\frac{1}{2}} \bm{Y}_{it}^\top 
  ))
\end{aligned}
\end{equation} 
where the last line follows from the tangent-space  
parametrization (\ref{parametrization}) of  $\bm{\Psi}_{i}$. 
Equation 
(\ref{likelihood}) indicates that the 
  likelihood is 
 in the form of a Gaussian likelihood of transformed responses, 
\begin{equation} \label{likelihood2}
 \bm{\Gamma}^\top 
 \widebar{\bm{\Sigma}}^{-\frac{1}{2}} \bm{Y}_{it} \sim N(\bm{0},\ \exp(\phi_{\widebar{\bm{\Psi}}}(\bm{\Psi}_{i})) 
 \ =  N(\bm{0},\ \exp(\mbox{diag}(\tilde{\bm{B}}\bm{x}_i + \tilde{\bm{z}}_i)) 
\end{equation}
 and no attempt will be made to estimate the 
  parameters 
$\bm{L}_i$ in (\ref{lfm}) 
unrelated with $\bm{x}_i$. 

We specify the prior  
$p(\bm{\Gamma}, \bm{\Psi}, \tilde{\bm{B}}, \bm{\Omega}) 
 = p(\bm{\Gamma}, \tilde{\bm{B}}, \bm{\Omega}) p( \bm{\Psi} | \bm{\Gamma},  \tilde{\bm{B}}, \bm{\Omega})  
$  in (\ref{posterior}) 
 as 
 \begin{equation}  \label{prior}
\begin{aligned}
  \propto \ 
  p(\bm{\Gamma}) p(\tilde{\bm{B}}) p(\bm{\Omega})
\exp\left\{
 -  \frac{1}{2} \sum_{i=1}^n (\vec{\phi}_{\widebar{\bm{\Psi}}}(\bm{\Psi}_{i}) - \tilde{\bm{B}}\bm{x}_i)^\top \bm{\Omega}^{-1} 
 (\vec{\phi}_{\widebar{\bm{\Psi}}}(\bm{\Psi}_{i}) -\tilde{\bm{B}} \bm{x}_i)  
 -\frac{n}{2}
 \log |\bm{\Omega} | 
 \right\}, 
\end{aligned}
\end{equation} 
 using independent priors $p(\bm{\Gamma}, \tilde{\bm{B}}, \bm{\Omega})  = p(\bm{\Gamma}) p(\tilde{\bm{B}}) p(\bm{\Omega})$ 
 and a conditional prior on 
 $\bm{\Psi} = \{ \bm{\Psi}_i \}_{i=1}^n$ 
given $(\bm{\Gamma}, \tilde{\bm{B}}, \bm{\Omega})$  based on 
$\phi_{\widebar{\bm{\Psi}}}(\bm{\Psi}_{i}) = \mbox{diag}(\tilde{\bm{B}}\bm{x}_i + \tilde{\bm{z}}_i)$.  
In (\ref{prior}), 
 $\vec{\phi}_{\widebar{\bm{\Psi}}}(\bm{\Psi}_{i}) \in \mathbb{R}^d$ 
denotes the vector of the diagonal elements of $\phi_{\widebar{\bm{\Psi}}}(\bm{\Psi}_{i}) \in \mathbb{R}^{d \times d}$. 
For $\tilde{\bm{B}} \in \mathbb{R}^{d \times q}$,   
 we use a mean zero matrix Gaussian prior with element-wise standard deviation 
$\sigma_{\tilde{B}_{jk}} > 0$.  
   For 
 $\bm{\Omega} \in \mbox{Sym}_d^{+}$, which we decompose into 
 $\mbox{diag}(\bm{\omega}) \tilde{\bm{\Omega}} \mbox{diag}(\bm{\omega})$, 
 we use 
 an unit-scale half-Cauchy distribution \citep{Gelman2006, Polson2012} on each element of the standard deviation vector $\bm{\omega} \in \mathbb{R}^d$
 (allowing for  the possibility of extreme values) 
 and a Lewandowski-Kurowicka-Joe (LKJ)  prior   \citep{LKJ2009} on the correlation matrix $\tilde{\bm{\Omega}}$   
  with  hyperparameter $\eta > 0$  (specifying  
 the amount of expected prior correlations). 
 For $\bm{\Gamma} \in \mathbb{R}^{p \times d}$, 
 we use a matrix angular central Gaussian (MACG) \citep[][]{Chikuse1990, Jupp1999}  with  hyperparameter $\bm{\Phi} \in \mbox{Sym}_p^{+}$. 
 An orthonormal random matrix 
$\bm{\Gamma}$ is said to be distributed as a MACG (with parameter $\bm{\Phi}$) if $\bm{\Gamma} \overset{d}{=} 
 \bm{U} (\bm{U}^\top\bm{U})^{-1/2}$, 
where $\bm{U}  \in \mathbb{R}^{p \times d}$ follows a $p \times d$ matrix normal distribution, 
whose density is 
\begin{equation} \label{U.distr}
f_{\bm{U}}(\bm{U}) = (2\pi)^{-pd/2} |\bm{\Phi}|^{-d/2} \exp(\mbox{tr}(- \bm{U}^\top \bm{\Phi}^{-1} \bm{U}/2)). 
\end{equation}
If the row covariance  $\bm{\Phi} = \bm{I}_{p}$, 
then the prior on $\bm{U}$ encodes no spatial information. 
In our illustrations, 
we employed flat priors on $\bm{\Gamma}$ and the correlation matrix $\tilde{\bm{\Omega}}$ 
(with $\bm{\Phi} = \bm{I}_{p}$ and $\eta = 1$, respectively), 
 and  weakly informative priors on $\tilde{\bm{B}}$, 
  using $\sigma_{\tilde{B}_{jk}}^2 = 2.5^2$.

\subsubsection{Posterior computation via polar expansion}

A Markov chain Monte Carlo (MCMC) sampling for $\bm{\Gamma}$ from the posterior  (\ref{posterior})  is challenging 
due to the restriction that $\bm{\Gamma}$ is in a Stiefel manifold. 
We will use polar expansion to transform the orthonormal parameter $\bm{\Gamma}$ 
 to an unconstrained object $(\bm{U})$ to work around this restriction.  
 Generally, 
``parameter expansion'' of a statistical model  
refers to methods which expand the parameter space by introducing redundant working 
parameters for computational purposes 
 \citep{Jauch2021}.    
By polar decomposition \citep{Higham1986}, 
any arbitrary matrix $ \bm{U} \in \mathbb{R}^{p \times d}$  
can be  
decomposed into two components, 
\begin{equation} \label{polar} 
\bm{U} = \bm{\Gamma}_{\bm{U}} \bm{S}_{\bm{U}},
\end{equation}  
where the first component $\bm{\Gamma}_{\bm{U}} := \bm{U}  (\bm{U}^\top \bm{U})^{-1/2} \in \mathbb{R}^{p \times d}$ is an orthonormal (rotation) matrix, 
and the second 
$\bm{S}_{\bm{U}} := (\bm{U}^\top \bm{U})^{1/2} \in \mathbb{R}^{d \times d}$ is a symmetric nonnegative (stretch tensor) matrix.

Using a MACG prior on $\bm{\Gamma}$ 
  with prior on $\bm{U}$ in (\ref{U.distr}) 
allows for posterior inference on 
   $\bm{U}$ (rather than directly on $\bm{\Gamma}$).  
By employing the polar expansion of $\bm{\Gamma}_{\bm{U}}$  
  to $\bm{U}$ in (\ref{polar}),  
we 
 ``parameter expand'' 
an orthonormal $\bm{\Gamma}_{\bm{U}}$ 
 to an unconstrained $\bm{U}$. 
 This expanded parameter maintains the same model likelihood 
 $p(\mathcal{D} | \bm{\Gamma}_{\bm{U}}, \bm{\Psi}, \tilde{\bm{B}},  \bm{\Omega})$ as in (\ref{likelihood}). 
  However, 
  the 
  prior 
 $p(\bm{\Gamma}_{\bm{U}}, \bm{\Psi}, \tilde{\bm{B}}, \bm{\Omega})$ 
 in (\ref{prior}) 
 expands to 
 $p(\bm{U}, \bm{\Psi}, \tilde{\bm{B}}, \bm{\Omega})$ 
 under 
 parametrization 
 (\ref{polar}), 
 leading to the corresponding posterior expansion from 
 $p(\bm{\Gamma}_{\bm{U}}, \bm{\Psi}, \tilde{\bm{B}}, \bm{\Omega} |\mathcal{D})$ in (\ref{posterior})  
 to 
 $p(\bm{U}, \bm{\Psi}, \tilde{\bm{B}}, \bm{\Omega} |\mathcal{D})$. 
Using MCMC,  
we first approximate samples from the expanded posterior $p(\bm{U}, \bm{\Psi}, \tilde{\bm{B}}, \bm{\Omega} |\mathcal{D})$,  
then conduct the polar decomposition (\ref{polar}) 
 to obtain the samples from the posterior of $\bm{\Gamma}_{\bm{U}}$, which can be verified  via a change of variable 
from $\bm{U}$ to 
 $\bm{\Gamma}_{\bm{U}}$.  
Specifically, given a Markov chain $\{ \bm{U}_s, \bm{\Psi}_s, \tilde{\bm{B}}_s, \bm{\Omega}_s \}$ 
with a stationary distribution 
proportional to $p(\bm{U}, \bm{\Psi}, \tilde{\bm{B}}, \bm{\Omega} |\mathcal{D})$, 
we approximate 
the  posterior 
of $\bm{\Gamma}$ by 
 $\{ \bm{\Gamma}_s \}$ 
  where 
 $\bm{\Gamma}_s =  \bm{U}_s (\bm{U}_s^\top \bm{U}_s)^{-1/2}$ for each $s$, 
 yielding 
 approximate 
 samples from $p(\bm{\Gamma}, \bm{\Psi}, \tilde{\bm{B}}, \bm{\Omega} |\mathcal{D})$. 
 
In this paper, 
approximate the posterior distribution of parameters  
 $(\bm{U}, \bm{\Psi}, \tilde{\bm{B}},  \bm{\Omega})$ 
using an adaptive Hamiltonian Monte Carlo (HMC) sampler 
 \citep{Neal2011} 
with automatic differentiation and adaptive tuning, 
 implemented in Stan \citep{Stan}. Consequently, 
we obtain HMC posterior samples of $(\bm{\Gamma}, \bm{\Psi}, \bm{B},  \bm{\Omega})$. 
The mapping between $\bm{B}$ and $\tilde{\bm{B}}$ is given in Supplementary Materials S1. 
As in any PCA-type analysis, there is a sign non-identifiability of $\bm{\Gamma}$; the non-identifiability of matrix $\bm{\Gamma}$ up to random sign changes for each component. That is, the component vector 
$\bm{\gamma}^{(k)}$ and $-\bm{\gamma}^{(k)}$ correspond to the same direction. 
 We can align the posterior samples 
 $\{\bm{\gamma}_s^{(k)}\}$. 
   For the first post-warmup sample $\bm{\gamma}_{1}^{(k)}$,   let $j_1 = \argmax_j (|\gamma_{j,1}^{(k)}|)$.  
 For $s \ge 2$, we compared the sign of  $\gamma_{j_1,s}^{(k)}$ with that of $\gamma_{j_1,1}^{(k)}$,  
    and if the signs disagreed, we multiplied  $\bm{\gamma}_s^{(k)}$  by -1.  The aligned $\bm{\gamma}_{s}^{(k)}$'s were used to construct the credible intervals of $\bm{\gamma}^{(k)}$.  
In Sections~\ref{sec.simulation} and ~\ref{sec.application}, 
we employed a burn-in of 700 steps, during which Stan optimizes tuning parameters for the HMC sampler. After burn-in, we ran HMC for an additional 1300 steps to generate 1300 post-warmup samples. Convergence was assessed by examining traceplots of random parameter subsets.

Unlike ICA, where the order of the extracted components is relatively arbitrary, 
the components 
 $\bm{\gamma}^{(k)\top} \bm{Y}_{it}$ $(k=1,\ldots, d)$  
 in (\ref{lfm})
  specified by $\bm{\Gamma} =  [\bm{\gamma}^{(1)}, \cdots, \bm{\gamma}^{(d)}] \in \mathbb{R}^{p \times d}$ 
can be ranked based on the sample variance of the expected log-variance $E[\log \check{\Psi}_i^{(k)} | \mathcal{D}]$ they explain 
across observations 
$i=1,\ldots,n$,  
where 
 $\log \check{\Psi}_i^{(k)} = \bm{x}_i^\top \bm{\beta}^{(k)}$ $(k=1,\ldots,d)$; 
 here we 
 exclude  subject-level random effects $z_{i}^{(k)} \in \mathbb{R}$ 
 to quantify only covariate-associated heteroscedasticity. 
Specifically, we 
sort the  $d$  estimated  
components  
 in decreasing order of the magnitude of the sample variance  
 $V^{(k)} = 
 \sum_{i=1}^n \left\{ E[\log \check{\Psi}_i^{(k)} | \mathcal{D}] - \frac{1}{n}\sum_{i=1}^n E[\log \check{\Psi}_i^{(k)} | \mathcal{D}] \right\}^2$ 
  $(k=1,\ldots, d)$  
   of the expected  log-variance  $E[\log \check{\Psi}_i^{(k)} | \mathcal{D}]$ attributable to $\bm{x}_i$.

\subsubsection{Determination of the number $d$ of the components}
 
We propose to use a selection criterion based on the Watanabe-Akaike Information Criterion (WAIC) \citep{Watanabe2010}  
which can be used to estimate the expected log posterior. 
Given a fixed $d$, we compute the 
log pointwise
 predictive 
density  (LPPD) 
of the dimension reduced model, 
penalized by the WAIC 
effective degrees of freedom, $\hat{r}_{waic}$ (e.g., \cite{Gelman2014}).  
Specifically, we select the dimensionality $d$ 
of the covariate-assisted outcome projection, 
which maximizes the expected \textit{deviance} between two models in the projected outcome space: 
one incorporating covariate-explained heteroscedasticity  
$\bm{\Gamma}^{\top} \bm{Y}_{it}  \sim N(\bm{0}, \check{\bm{\Psi}}_i = \exp(\mbox{diag}(\bm{B}\bm{x}_i)))$, 
 and the other 
 without heteroscedasticity 
$\bm{\Gamma}^{\top} \bm{Y}_{it}  \sim N(\bm{0}, \widebar{\bm{\Psi}} = \bm{\Gamma}^\top \widebar{\bm{\Sigma}} \bm{\Gamma})$. 
The expected 
deviance (scaled by $-2$) 
 is estimated by 
\begin{equation} \label{WAIC}
-2 \ \sum_{i=1}^n \sum_{t=1}^{T_i} 
\frac{1}{S}\sum_{s=1}^S 
\log R^{(s)}
+ 2 \hat{r}_{waic},
\end{equation} 
where
$\hat{r}_{waic} = \sum_{i=1}^n \sum_{t=1}^{T_i}   \frac{1}{S}\sum_{s=1}^S \left( 
\log R^{(s)}   
- \frac{1}{S}\sum_{s=1}^S  \log R^{(s)}  
\right)^2,$  
in which 
$R^{(s)} = \frac{p(\bm{\Gamma}^{(s)\top} \bm{Y}_{it} |\check{\bm{\Psi}}_i^{(s)})}{p(\bm{\Gamma}^{(s)\top} \bm{Y}_{it} |\widebar{\bm{\Psi}}^{(s)})}$,
i.e., the posterior ratio of the two models with vs. without covariate-explained 
heteroscedasticity, 
computed using the MCMC 
posterior parameter samples $(s=1,\ldots,S)$. 
If the covariates
 $\bm{x}_i$ 
 are predictive of the covariances 
 $\bm{\Gamma}\bm{\Sigma}_i \bm{\Gamma}$
 along all  PDCs 
 $\bm{\Gamma} \in \mathbb{R}^{p \times d}$ of rank $d$,  
 then 
the corresponding expected log posterior, $E\left[\log p(\bm{\Gamma}^{\top} \bm{Y}_{it} |\check{\bm{\Psi}}_i)  \right]$, 
will be large.   
However,  for a too large rank $d$,
the covariates may not predict the covariances 
  $\bm{\Gamma}\bm{\Sigma}_i \bm{\Gamma}$ in all posited directions $\bm{\Gamma}$, 
leading to a smaller expected log posterior ratio,  
$E\left[\log \frac{p(\bm{\Gamma}^{\top} \bm{Y}_{it} |\check{\bm{\Psi}}_i)}{p(\bm{\Gamma}^{\top} \bm{Y}_{it} |\widebar{\bm{\Psi}})} \right] 
= E\left[\log p(\bm{\Gamma}^{\top} \bm{Y}_{it} |\check{\bm{\Psi}}_i)  \right] - E\left[\log p(\bm{\Gamma}^{\top} \bm{Y}_{it} |\widebar{\bm{\Psi}})  \right]$, 
compared to that with the 
optimal projected outcome dimension $d$. 
 Considering the ratio  is crucial for making this   criterion comparable across different $d$'s, 
  and we select $d$ that minimizes this expected deviance. 
In Supplementary Materials S2, we demonstrate the validity of this criterion in selecting the correct number of covariate-relevant heteroscedasticity components.

\section{Simulation illustration} \label{sec.simulation}

\subsection{Simulation setup} \label{sec.setup}
 
 For each unit (subject) $i$, we simulate a set of outcome signals 
$\bm{Y}_{it} \in \mathbb{R}^p$ $(t=1,\ldots,T_i)$  $(i=1,\ldots,n)$ 
from a Gaussian distribution with mean zero 
and $p \times p$ unit-specific covariance $\bm{\Sigma}_{i}$. 
 We vary $n \in \{100, 200, 300, 400\}$, $T_i \in \{10, 20,30\}$, 
 and $p \in \{10, 20\}$. We use  model (\ref{cov.model}) 
to generate $\bm{\Sigma}_i \in \mathbb{R}^{p \times p}$,  
 where the core SPD $\bm{\Psi}_i = \exp(\mbox{diag}(\bm{B} \bm{x}_i + \bm{z}_i)) \in \mbox{Sym}_d^{+}$ with $d=2$,
 where  $\bm{x}_i 
= (1, x_{i1}, x_{i2}, x_{i3}, x_{i4})^\top \in \mathbb{R}^q$, is defined based on the subject-level linear predictors 
$\bm{B} \bm{x}_i + \bm{z}_i$, 
\begin{equation*}  \label{sim.lp}
\bm{B} \bm{x}_i + \bm{z}_i = \begin{pmatrix}  
0.1 & 0.4 & -0.5  & 0.5 & -0.5 \\
0.1 & -0.3 & 0.4 & -0.4 & 0.4 \\ 
 \end{pmatrix} 
 \begin{pmatrix} 
1 \\
x_{i1} \\
x_{i2} \\
x_{i3} \\
x_{i4}
 \end{pmatrix} 
 + \bm{z}_i   
 = 
 \bm{\beta}_0   + 
     \begin{pmatrix}  
      (x_{i1} \ x_{i2} \ x_{i3} \ x_{i4}) \bm{\beta}^{(1)}    \\
 (x_{i1} \ x_{i2} \ x_{i3} \ x_{i4}) \bm{\beta}^{(2)}     \\
    \end{pmatrix} 
    + \bm{z}_i
\end{equation*}  
of dimension $d=2$, 
where $\bm{\beta}_0 = (0.1, 0.1)^\top \in \mathbb{R}^2$ is the intercept vector,  
 $\bm{\beta}^{(1)} = (0.4,  -0.5, 0.5, -0.5)^\top$  and $\bm{\beta}^{(2)} = (-0.3,  0.4, -0.4,  0.4)^\top$ 
are the regression coefficients for 
$(x_{i1}, x_{i2}, x_{i3}, x_{i4})^\top \in \mathbb{R}^{q-1}$. We generate covariates  $x_{i1} \overset{iid}{\sim} \mbox{Bernoulli}(0.5)$ 
and $x_{i2},x_{i3}, x_{i4} \overset{iid}{\sim} N(0, 1^2)$, 
and the 
 subject-specific random effects 
$\bm{z}_i \overset{iid}{\sim}  N(\bm{0}, \bm{\Omega})$,   
where  
  $\bm{\Omega} = 
   \begin{pmatrix}  
\omega_{11} & \omega_{12}  \\
\omega_{12} & \omega_{22} \\ 
 \end{pmatrix}
 =
 \begin{pmatrix}  
0.5^2 &  0.1  \\
0.1 & 0.5^2 \\ 
 \end{pmatrix}
$, to define $\bm{\Psi}_i$. 

For each simulation run, we 
use the von Mises-Fisher distribution to 
randomly generate 
an orthonormal basis matrix $[ \bm{\Gamma}, \bm{L}] \in \mathbb{R}^{p \times p}$ for  
$\bm{Y}_{it} \in \mathbb{R}^p$,  
and its  subcomponent $ \bm{L} \in \mathbb{R}^{ p \times (p-d)}$ is 
further transformed by 
subject-specific orthonormal matrices  
$\bm{A}_i \in \mathbb{R}^{(p-d) \times (p-d)}$, each   randomly  generated from the von Mises-Fisher distribution. 
Then, 
the 
 ``noise''  covariance 
components  
 $\bm{L}_i  \bm{L}_i^\top 
= 
\bm{L} \bm{A}_i \exp\big(\mbox{diag}(\bm{\epsilon}_i)\big) \bm{A}_i^\top \bm{L}^\top  
\in \mathbb{R}^{p \times p}$ 
are specified by generating $\bm{\epsilon}_i \in \mathbb{R}^{p-d}$ with 
each element  $\epsilon_{ij} \overset{iid}{\sim} N(0, 0.5^2)$, 
whereas $\bm{\Gamma} \exp\big( \mbox{diag}(\bm{B} \bm{x}_i + \bm{z}_i)\big)  \bm{\Gamma}^\top \in \mathbb{R}^{p \times p}$ 
specify the ``signal'' 
components. 
For each simulation run,  we compute the base covariance $\widebar{\bm{\Sigma}}$ 
that we use for tangent-space parametrization of model (\ref{cov.model}) 
as the sample marginal covariance on the training sample. 
  
  To investigate the robustness of the method against model misspecification, 
 we further 
consider the case where there are no common eigenvectors 
$\bm{\Gamma}$ across subjects. 
We consider 
 subject-level random perturbation 
 using the subject-level rotation matrices 
 $\bm{R}(\theta_i) = 
 \begin{pmatrix} 
 \cos(\theta_i) & -\sin(\theta_i)  \\
 \sin(\theta_i) &  \cos(\theta_i)  \\ 
 \end{pmatrix}$  
 with 
 random angles  
 $\theta_i \overset{iid}{\sim}  \mbox{Unif}[-\pi/10,\pi/10]$ $(i=1,\ldots,n)$, 
and use 
 $\bm{\Gamma} \bm{R}(\theta_i) \in \mathbb{R}^{p \times d}$ $(i=1,\ldots,n)$ in place of 
  $\bm{\Gamma} $ 
  in generating the responses in (\ref{lfm}), 
  referred to as ``model misspecification'' cases.

 \subsection{Evaluation metric}

We run the simulation 50 times. 
For each simulation run, we compute, as evaluation metrics, 
the absolute cosine similarity 
  $1- |\langle \widehat{\bm{\gamma}}^{(k)}, \bm{\gamma}^{(k)} \rangle |$ 
  for the loading coefficient vectors 
  (where a value close to $0$ indicates the proximity) 
  and the root mean squared error (RMSE) $\lVert \widehat{\bm{\beta}}^{(k)} -  \bm{\beta}^{(k)} \rVert/\sqrt{4}$ $(k=1,2)$  for the regression coefficient vectors, 
as well as the 
RMSE  for the elements of the random effect covariance matrix $\bm{\Omega}$,  $\lVert (\widehat{\omega}_{11},  \widehat{\omega}_{12}, \widehat{\omega}_{22})^\top - (\omega_{11},  \omega_{12}, \omega_{22})^\top \rVert/\sqrt{3}$, 
  where the notation $\widehat{\cdot}$ represents the posterior mean of $\cdot$. 
  While we conduct 
  the model estimation 
using the tangent space parameterization (\ref{lp}) 
with $\tilde{\bm{B}}$,   the results 
  are mapped 
  to the original parametrization with  
  $\bm{B}$   in (\ref{cov.model}). 
This approximately amounts to 
  shifting the intercept vector $\bm{\beta}_0 := 
  (\beta_0^{(1)}, \beta_0^{(2)})^\top \in \mathbb{R}^2$ by 
  the diagonal elements of 
  $\log(\bm{\Gamma}^\top \widebar{\bm{\Sigma}}^{-1} \bm{\Gamma}) \in \mathbb{R}^{2 \times 2}$ (see Supplementary Materials S1).   
We  
report the estimation performance 
 for $\bm{\beta}_0$ 
 by reporting RMSE $\lVert \widehat{\bm{\beta}}_0 -  \bm{\beta}_0 \rVert/\sqrt{2}$, 
under the original parametrization with $\bm{B}$.  
Additionally, to assess whether the constructed credible intervals provide reasonably correct coverage for the true values of the parameters, 
we evaluate the posterior credible intervals of the model parameters 
($\bm{\gamma}^{(k)}$, $ \bm{\beta}^{(k)}$, $\bm{\Omega}$) 
with respect to the
frequentist's coverage proportion. 
 Specifically, for each simulation run, 
 we  estimate the posterior distribution of the parameters  and 
calculate the  95\% posterior
 credible intervals for the parameters, 
and then evaluate  
   how often the credible intervals contain the true parameter values. 
   We used a random initialization of the Markov chains in our posterior sampling.  

   \vspace{-0.05 in}

\subsection{Simulation results}

\begin{figure}[H] 
\begin{center}
\begin{tabular}{c}  
\includegraphics[width=5.5  in, height = 3.5 in]{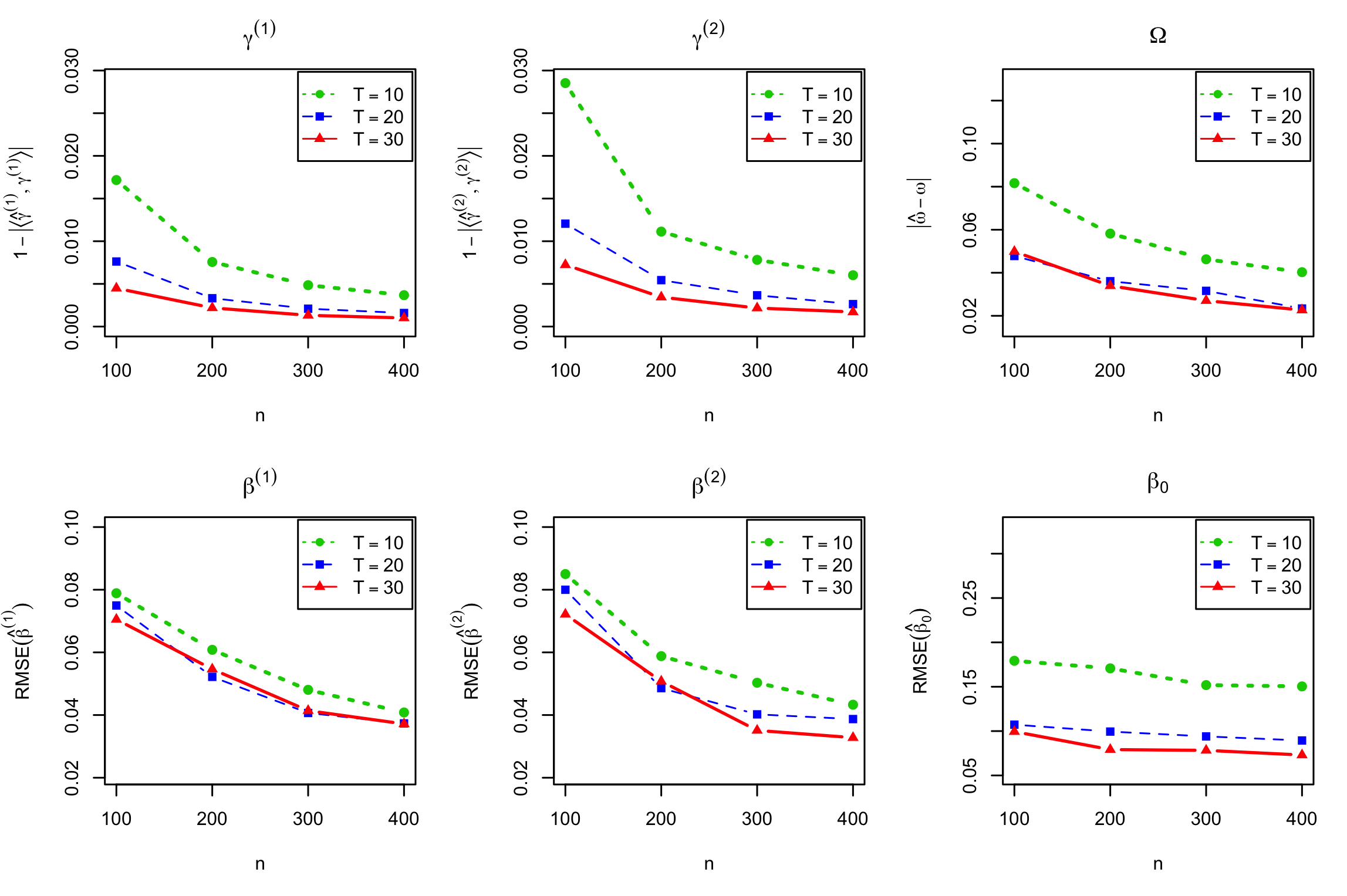}  
\end{tabular}  
\end{center}
  \vspace{-0.35 in}
\caption{ 
The model parameter estimation performance for $p=20$ case, 
for the loading coefficient vectors $\bm{\gamma}^{(k)}$ $(k=1,2)$, 
elements of the random effect covariance matrix $\bm{\Omega}$, 
regression coefficients $\bm{\beta}^{(k)}$ $(k=1,2)$, and  
  intercept $\bm{\beta}_0$, 
averaged across 
 $50$ simulation replications, 
with varying $n \in \{100, 200, 300, 400\}$ and $T \in \{10, 20, 30\}$.
} \label{fig.sim.p20}
\end{figure}

    \vspace{-0.1 in}

In Figure~\ref{fig.sim.p20}, 
as sample sizes ($n, T)$ increase, 
the estimation performance tends to improve overall. 
Particularly when the sample sizes are relatively small (e.g., $n=100, T=10$), 
 the 
 improvement 
tends to depends on 
the magnitude of the covariate effects on the outcome projection component, 
as  
 performance for parameters 
 for the first 
  component 
($\bm{\gamma}^{(1)}$ and $\bm{\beta}^{(1)}$) 
tends to be slightly better than those for the second components  
($\bm{\gamma}^{(2)}$ and $\bm{\beta}^{(2)}$), 
reflecting stronger covariate effects on 
 the first projection component. 
The number of subjects ($n$) and time points ($T$) both influence performance; increasing $T$ 
enhances estimation by providing more subject-level information for accurate estimates of subject-specific random effects and their covariance $\bm{\Omega}$, and accordingly population-level parameters  $\bm{\gamma}^{(k)}$ and $\bm{\beta}^{(k)}$. 
 The $p = 10$ cases reported in Supplementary Materials S3 show qualitative similar results to those for the $p = 20$ cases.

    \vspace{-0.05 in}
\begin{table} [H]
\begin{centering}
\begin{tabular}{l | l | ccccc|cccccccccc cccc|ccc}  
\toprule
   &   & \multicolumn{5}{c}{$p=10$} &   \multicolumn{5}{|c}{$p=20$} \\
\toprule
$n$    & $T$ &	$\bm{\gamma}^{(1)}$   & $\bm{\gamma}^{(2)}$ & $\bm{\beta}^{(1)}$ &  $\bm{\beta}^{(2)}$ &  $\bm{\Omega}$  &	$\bm{\gamma}^{(1)}$   & $\bm{\gamma}^{(2)}$ & $\bm{\beta}^{(1)}$ &  $\bm{\beta}^{(2)}$ &  $\bm{\Omega}$   \\
\toprule
100   & 10   \quad  & 0.89 &0.90 &0.93 &0.90 &0.91 &0.86 &0.86 &0.86 &0.84 & 0.88   \\ 
         &  20  \quad  & 0.85 &0.85 &0.91 &0.87 &0.93 &0.90 &0.89 &0.92 &0.87  &0.94 \\ 
         &  30  \quad  &  0.88 &0.87& 0.91& 0.90 &0.91 &0.87& 0.89& 0.88 &0.88 & 0.88 \\ 
\hline
200   & 10   \quad &  0.90 &0.88 &0.95 &0.92 &0.94 &0.90 &0.93 &0.96 &0.94 & 0.89    \\ 
         &  20  \quad  & 0.92 &0.91 &0.97 &0.92& 0.93 &0.91 &0.92 &0.93& 0.94  &0.93 \\ 
         &  30  \quad  &  0.89 &0.89& 0.96 &0.88& 0.94& 0.89 &0.89& 0.90& 0.89  &0.89 \\ 
\hline
300   & 10   \quad & 0.88 &0.88 &0.90 &0.90 &0.88 &0.90& 0.91 &0.96 &0.92  &0.89 \\ 
         &  20  \quad  & 0.90 &0.86 &0.90 &0.84 &0.91 &0.92& 0.90& 0.96 &0.92 & 0.90 \\ 
         &  30  \quad  & 0.91 &0.90 &0.91 &0.90& 0.91& 0.91 &0.91& 0.94 &0.92 & 0.91\\ 
\hline
400   & 10   \quad & 0.91 &0.89 &0.96 &0.92 &0.85& 0.90 &0.93 &0.93&0.92  &0.90 \\ 
         &  20  \quad  &  0.94 &0.91 &0.96 &0.96& 0.87 &0.92& 0.91 &0.94 &0.92  &0.93\\ 
         &  30  \quad  &  0.93 &0.92 &0.96 &0.95 &0.89 &0.93 &0.91 &0.92 &0.92  &0.92\\ 
\bottomrule
\end{tabular}
\caption{The proportion of time that 
95\% posterior credible intervals contain the true values of the projection loading vectors $\bm{\gamma}^{(k)}$ $(k=1,2)$, 
regression coefficients $\bm{\beta}^{(k)}$  $(k=1,2)$, and 
 elements of $\bm{\Omega}$, 
 averaged across $50$ simulation replications, 
with varying $n \in \{100, 200, 300, 400\}$ and $T \in \{10, 20, 30\}$. 
Coverage computed for each entry, 
 then averaged within components ($\bm{\gamma}^{(k)}$, $\bm{\beta}^{(k)}$ and $\bm{\Omega}$) and across the 
  simulation 
 replications (rounded to two significant digits). 
}
\label{tab.coverage.sim}
\end{centering}
\end{table}

\vspace{-0.1 in}
 
In terms of coverage probability, 
the results in Table~\ref{tab.coverage.sim}  for both $p=10$ and $20$ cases
 indicate that  the ``actual'' coverage probability is 
reasonably close to the ``nominal'' coverage probability of $0.95$, 
particularly 
with larger sample sizes 
 (e.g., $n=400$, $T=30$) for the regression coefficients $\bm{\beta}^{(k)}$. 
Overall, the results  in Table~\ref{tab.coverage.sim} suggest that 
the Bayesian credible intervals exhibit  
reasonable frequentist coverage, 
providing estimates of the parameter uncertainty that aligns with the desired coverage level. 
In Supplementary Materials S4, 
we further examine the model's performance under misspecification: 
1) when excluding the random effect component  
$\bm{z}_i$; 
and 
2) when there  are no  common ``signal'' eigenvectors across subjects. 
Without the random effect,  estimation performance remains comparable in terms of bias, 
but the coverage of  95\% credible intervals tends to underestimate uncertainties, particularly for the regression coefficients $\bm{\beta}^{(k)}$. 
The absence of common covariate-related eigenvectors introduces bias in estimating $\bm{\beta}^{(k)}$, leading to lower coverage levels of the credible intervals than nominal. 
The average computation time (on a MacBook running M3 Max with 96 GB unified memory) was about 0.8 hours (SD= 0.16) for obtaining 1300 posterior samples on $n=400$ subjects with $T=30$ time points and $p = 20$.

\section{Application} \label{sec.application}

In this section, 
we applied the Bayesian CAP regression to
 data from HCP. As in \cite{Seiler2017}, 
 we used 
  the 
  rs-fMRI data from HCP 820 subjects 
 and examined the associations between rs-fMRI and sleep duration. 
Each subject underwent 4 complete 15-minute sessions (with TR = 750ms, corresponding to 1200 time points per session for each subject), 
and 
each 15-minute run of each subject’s rfMRI data was preprocessed according to \cite{Smith2013}. 
We focused on the first session which is about a typical  duration for rs-fMRI studies. 
We also applied the proposed method to the other three sessions to examine the sensitivity and  reliability of this regression  
 (see Supplementary Materials S6, 
where the covariate-related FC exhibits a high level of consistency across all 4 scanning sessions, with the intra-cluster correlation coefficient value of 0.84, 0.72, 0.84 and 0.83, for the 4 identified network components in terms of the log-variance). 

\begin{figure}[H] 
\begin{center}
\begin{tabular}{c} 
\includegraphics[width=4.8  in, height = 1.8  in]{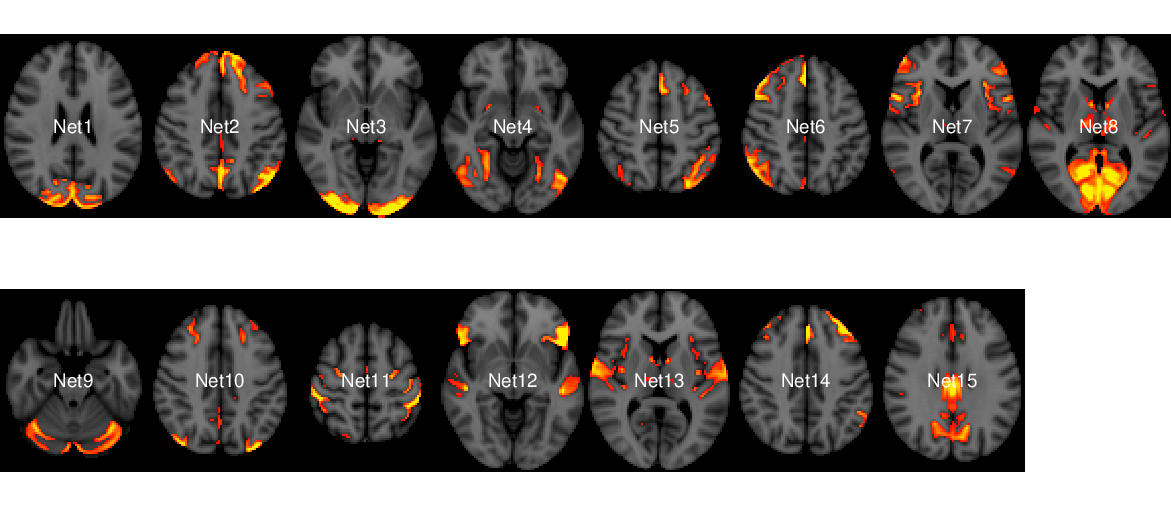}  
\end{tabular}  
\end{center}
  \vspace{-0.3 in}
\caption{15 independent components (ICs) from spatial group-ICA constituting a data-driven parcellation with 15 components (``network nodes''), provided by the HCP PTN dataset, represented at the most relevant axial slices in MNI152 space. 
According to \cite{Seiler2017}, 
these 
IC networks correspond to default network (Net15), 
cerebellum (Net9), visual areas (Net1, Net3, Net4 and Net8), 
cognition-language (Net2, Net5, Net10 and Net14), perception-somesthesis-pain (Net2, Net6, Net10 and Net14), 
sensorimotor (Net7 and Net11), executive control (Net12) and auditory (Net12 and Net13). 
} \label{fig1}
\end{figure}

We used a data-driven parcellation based on spatial ICA with $p=15$ components
 (i.e., using $p=15$ data-driven ``networks nodes''; see Figure~\ref{fig1} for their most relevant axial slices in MNI152 space) from the HCP 
PTN (Parcellation + Timeseries + Netmats) dataset,   
where each subject's rs-fMRI timeseries data were mapped onto the set of ICA maps  \citep{Filippini2009}. We refer to \cite{Smith2013} for details about preprocessing and the ICA time series computation. 
We conduct inference on the association between the FC over these IC network nodes  \citep{Smith2012}  and 
 sleep duration, gender and their interaction.

As in \cite{Seiler2017},  
 we classified the subjects into two groups: a group of 489 conventional sleepers (average sleep duration between 7 and 9 hours each night)  and  a group of
  241 short sleepers (average equal or less than 6 hours each night). 
This yielded a total of 730 participants to compare FC (over the IC networks in Figure~\ref{fig1}) between short and conventional sleepers. 
Since the time series are temporally correlated, 
we inferred the equivalent sample size of independent samples. 
We computed the effective sample size (ESS) defined by \cite{Kass1998},  
$
\mbox{ESS} = \underset{i \in \{1,\ldots,n\}, j \in \{1,\ldots,p\}}{\min} \left( \frac{T_i}{ 1 + 2 \sum_{s=1}^{\infty} \mbox{cor}(Y_{i1}^{(j)}, Y_{i,1+s}^{(j)}) } \right),
$
where $Y_{it}^{(j)}$ is the data at time $t$ of the $j$th network node for subject $i$, 
following a conservative approach taking the minimum over all $p$ components and $n$ subjects as the overall estimator. 
Based on the estimated ESS, we performed thinning of the observed timeseries data, 
subsampling $T_i = T = \mbox{ESS} = 34$ time points for each subject. 
The resulting outcome data, $\bm{Y}_{it}$ $(i=1,\ldots,n)$ $(t=1,\ldots,T)$, 
were then mean-removed per each subject (so that $\sum_{i=1}^T \bm{Y}_{it} = \bm{0} \in \mathbb{R}^{15}$ for each $i$), and we focused 
on the association between 
their covariances $\bm{\Sigma}_i$  and covariates.  

We used the  
WAIC criterion (\ref{WAIC}) to identify 
$d=4$ projection
 components. 
The models' WAIC values over the range of $d=1$ to $ 6$ 
were 
-227.9, 
-397.6
-520.4,
-602.7,
-573.4 
and 
-358.4, 
where the minimizer was the $d=4$ case. 
The parameters 
($\bm{\Gamma}$, $\bm{B}$ and $\bm{\Omega}$) 
  with $d=4$ 
 are summarized by their posterior means and 
 95\% credible intervals, reported in Supplementary Materials S4. 
  The expected value of 
  the log Deviation from Diagonality (DfD) was 0.60, 
suggesting a moderate departure from the diagonality of $\bm{\Psi}_i$ assumed in (\ref{lfm2}), but the deviation is not overly pronounced.

\begin{figure}[H]  
\includegraphics[width=6.4 in, height = 4.6in]{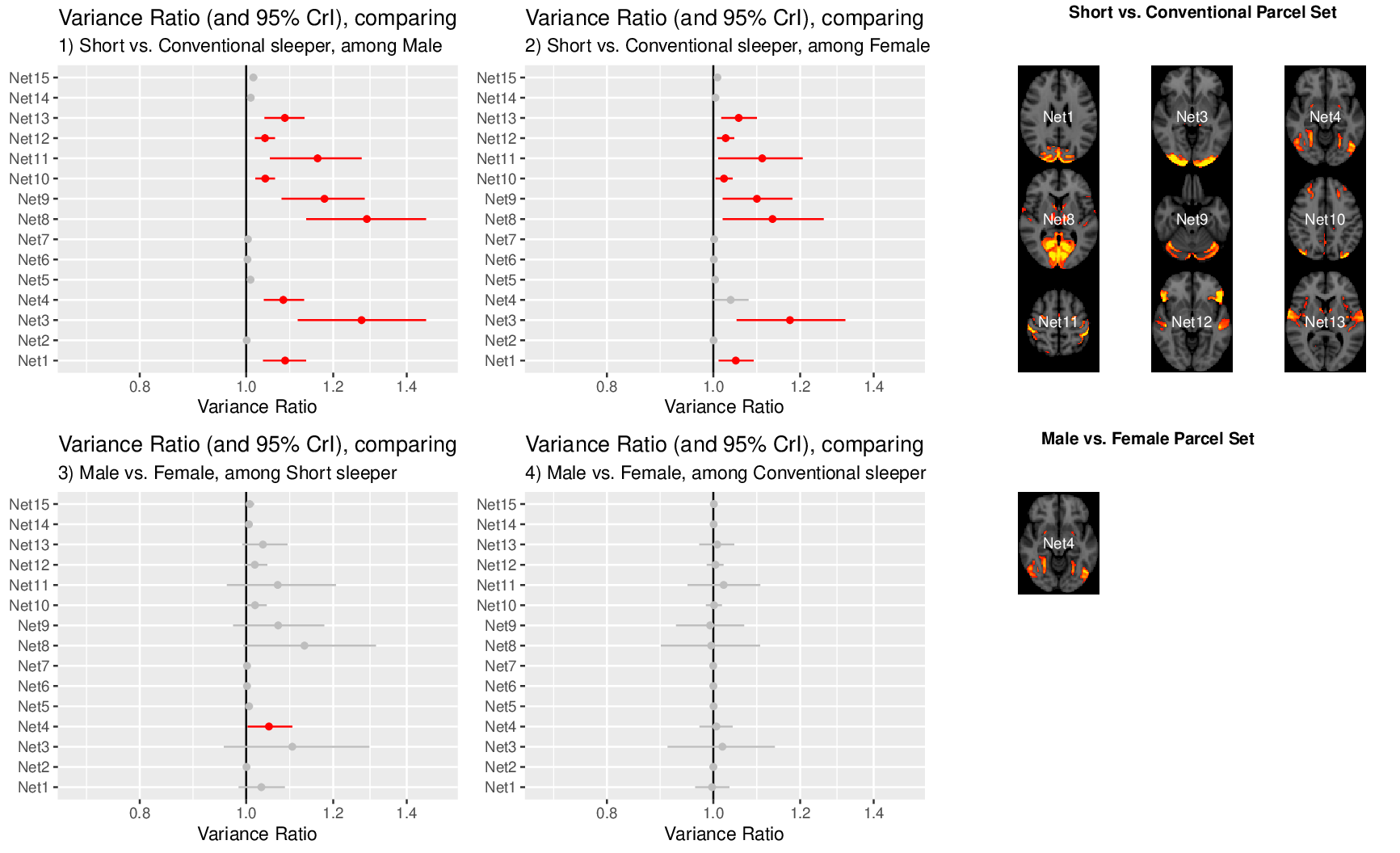}   \\  
  \vspace{-0.4 in}
\caption{
The response signals' variance ratio  (posterior means and 95\% credible intervals), 
corresponding to the four contrasts formed by the Gender-by-SleepDuration interaction.  
The 95\% credible intervals that do not include the variance ratio of 1 
are highlighted in red. The sets (``parcel sets'')  of network nodes  
whose signals' variances are expected to change in the same impact directions 
due to the corresponding contrasts 
are indicated in the last column panels,  
for the Short vs. Conventional sleeper contrasts in the top row, and 
 for the Male vs. Female contrasts in the bottom row. 
} \label{fig.VR}
\end{figure}

Under model (\ref{lfm}), 
for a linear contrast vector $\bm{\delta} \in \mathbb{R}^q$,  
we can define the  log  covariance ``contrast'' map due to a $\bm{\delta}$-change in the covariates $\bm{x} \in \mathbb{R}^q$, 
which 
corresponds to 
$\bm{\Gamma} 
\left( 
\mbox{diag}(\bm{B} \bm{\delta} )
\right)
\bm{\Gamma}^\top \in \mathbb{R}^{p \times p}$ (see Supplementary Materials S7), 
where  $\bm{B} \in \mathbb{R}^{d \times q}$ is 
the regression coefficient matrix  in 
(\ref{lfm2}).  
Specifically, the diagonal elements of this contrast 
  matrix 
$\bm{\Gamma} 
\left( 
\mbox{diag}(\bm{B} \bm{\delta}) 
\right)
\bm{\Gamma}^\top$ 
can be extracted  
and exponentiated. 
This represents   
the response signals' 
variance ratio (VR) 
corresponding to a $\bm{\delta}$-change in the 
covariates. 
For the four contrasts derived from  
the SleepDuration $\times$  Gender 
interaction,  the left two column panels in 
Figure~\ref{fig.VR} present the response signals' variance ratio, contrasting 
1) short vs. conventional sleeper among male; 
 2) short vs. conventional sleeper among female; 
 3) male vs. female among short sleeper; and 
 4) male vs. female among conventional sleeper. 
 
 In Figure~\ref{fig.VR}, the nodes, or ``parcels'',  
whose VR values were identified 
(based on 95\% credible intervals) 
to be significantly different from 1, 
were all with VR $>1$. 
The third column panels of Figure~\ref{fig.VR} 
indicate 
the nodes whose signals' variances are expected to change in the same direction, 
for the Short vs. Conventional sleeper contrasts in the top row panel, and 
 for the Male vs. Female contrasts in the bottom row panel.

\begin{figure}[H] 
\begin{center}
\begin{tabular}{c}  
\includegraphics[width=4.6 in, height = 2 in]{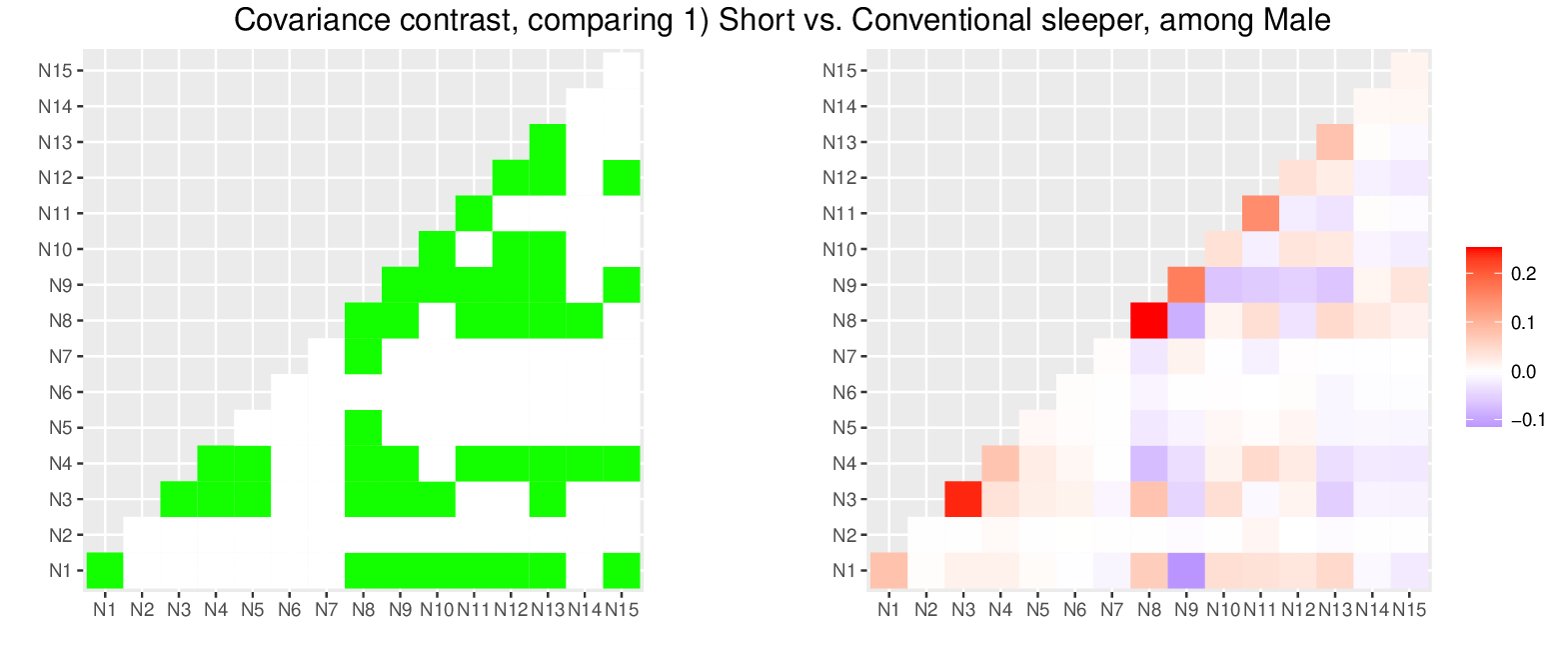}  \\
\includegraphics[width=4.6 in, height = 2 in]{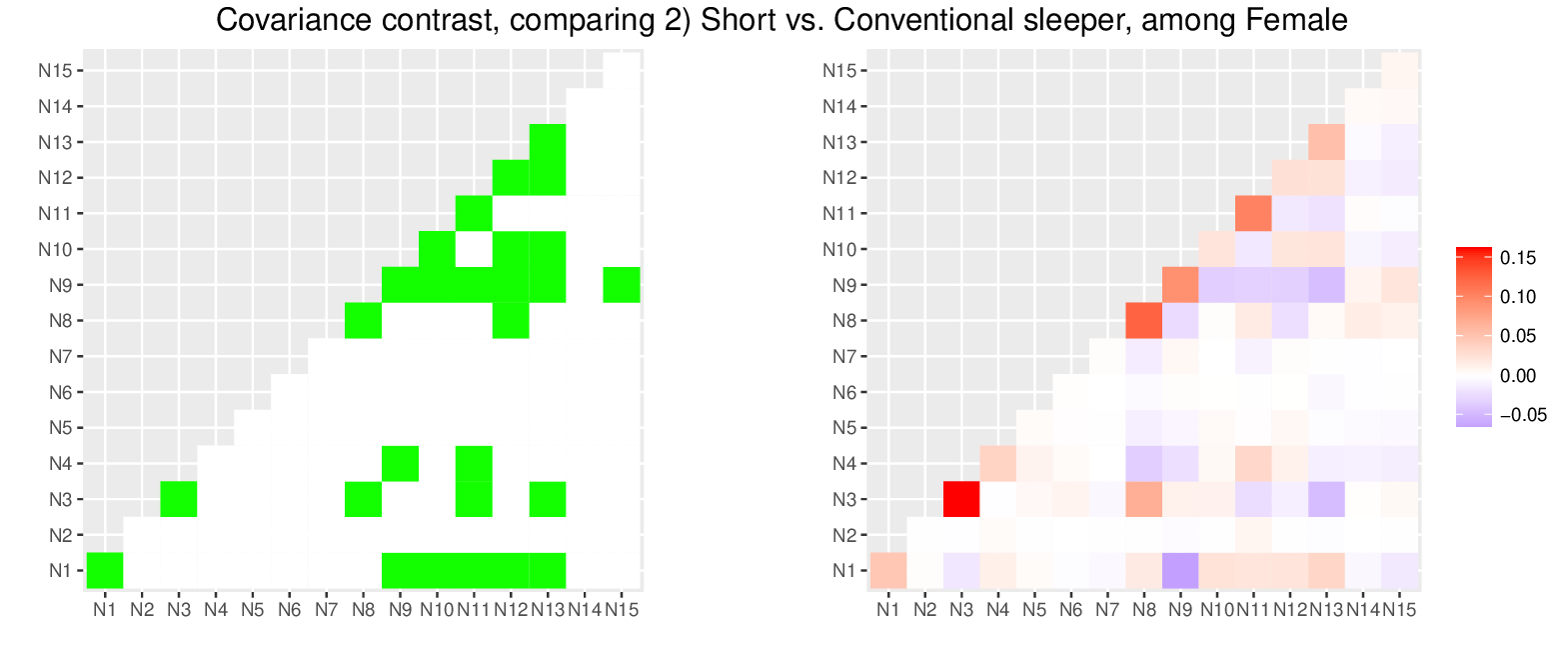}  \\
\includegraphics[width=4.6 in, height = 2 in]{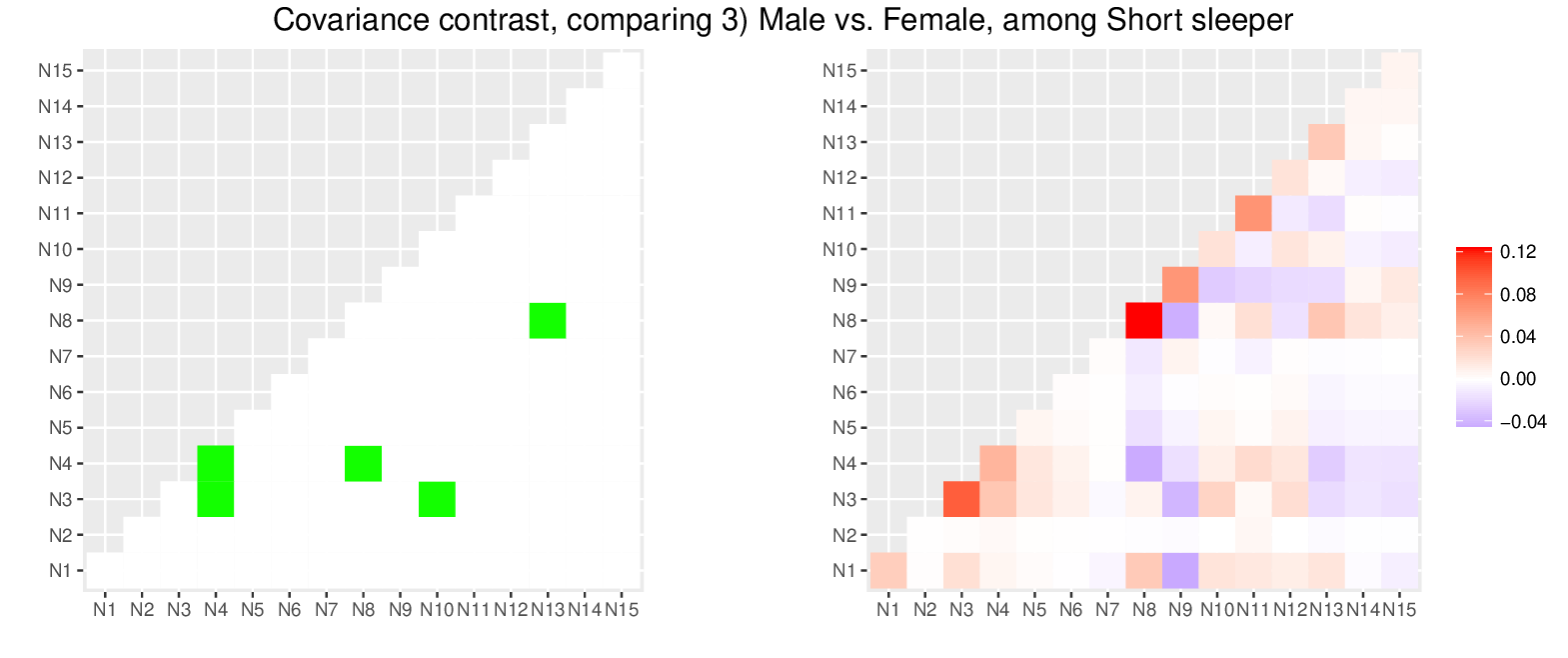}   \\
\includegraphics[width=4.6 in, height = 2 in]{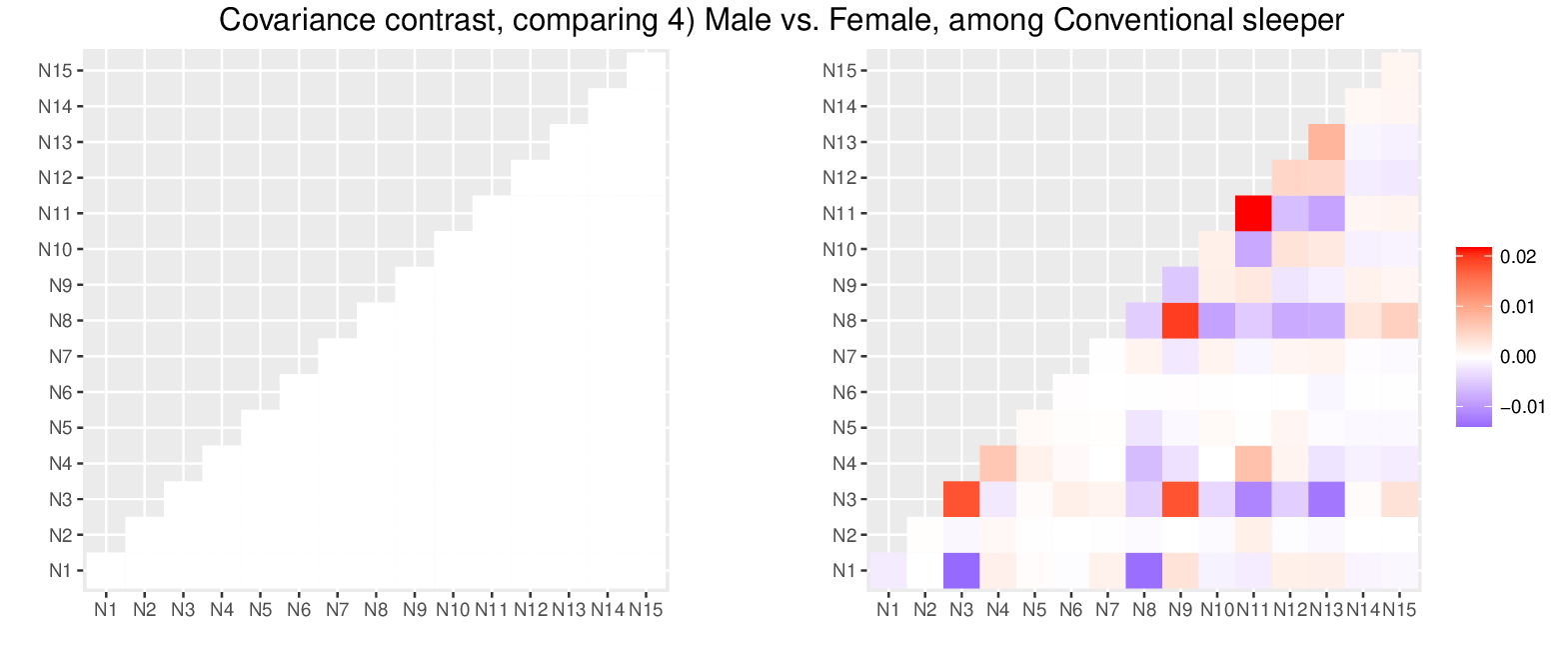}    
\end{tabular}  
\end{center}
  \vspace{-0.2 in}
\caption{
The statistical significance map (the left column panels) 
and the posterior mean (the right column panels) of the log covariance contrast 
$\bm{\Gamma} 
\left( 
\mbox{diag}(\bm{B} \bm{\delta} )
\right)
\bm{\Gamma}^\top$  
for each of the four covariate contrasts $\bm{\delta}$, derived from the SleepDuration $\times$ Gender interaction. 
} \label{fig.offdiag}
\end{figure}

For each $\bm{\delta}$ contrast, 
we 
 can infer the $\bm{\delta}$ contrasts' 
impact on the connectivity by 
95\% credible intervals 
on the $p(p+1)/2$ connectivity elements of the contrast matrix $\bm{\Gamma} \left( \mbox{diag}(\bm{B}\bm{\delta}) \right)\bm{\Gamma}^\top$. 
 The first column panels in Figure~\ref{fig.offdiag} display the   covariance elements 
identified to be significant (colored in green), 
 whereas the second column panels display the posterior mean of the matrix elements of $\bm{\Gamma} \left(\mbox{diag}(\bm{B}\bm{\delta})\right)\bm{\Gamma}^\top$, 
where each row panel corresponds to each $\bm{\delta}$ contrast in the covariates.  
 The results from the statistical significance maps in Figure~\ref{fig.offdiag} indicate that, 
 overall, there are more substantial connectivity differences between Short and Conventional sleepers (the first two row panels), 
 compared to the cases when we compare 
 Male vs. Female (the last two row panels), 
 and there were slightly more pronounced Short vs. Conventional sleepers differences among Males 
 (the first row panel) 
 than among Females  (the seond row panel).
While there were several identified connectivity differences between Male vs. Female among Short sleepers, 
there were no statistically significant Male vs. Female differences  
among Conventional sleepers. 
 
 One conventional method for analyzing group ICA data involves initially computing subject-level Pearson correlations between the ICs, which are then Fisher z-transformed. 
 This process is performed on $(p(p-1)/2=)$ $105$ pairs of correlations (calculated from $15$ ICs), 
while we conduct the element-wise log transformation  
 on the $p=15$ diagonal elements.  
A total of 120 element-wise linear regressions were then conducted  
on SleepDuration, Gender and their interaction, and  p-values were corrected for multiplicity using the Benjamini-Hochberg (BH) \citep{Benjamini} procedure to control the false discovery rate (FDR) at 0.05.  
The patterns of the connectivity differences, 
 implied by each $\bm{\delta}$-contrast, 
 from this mass-univariate approach 
 are presented in Supplementary Material S10,  
which were similar to the results from Bayesian CAP in Figure~\ref{fig.offdiag}. 
However, compared to the results from Bayesian CAP, 
far fewer statistically significant elements (13 vs. 77, out of 480 elements) were identified.   

While the CAP regression formulation of \cite{Zhao2021} 
 also alleviates the multiplicity issue and thus can improve statistical power, 
 inference is limited to the association between covariates and the projected outcome components, 
 making it challenging to interpret covariates' impacts 
 in measured ROIs directly.  
Therefore,  the approach is not directly comparable with the proposed approach here.    
In Supplementary Materials S8, 
we display the similarity (similarity between -1 and 1, with 0 indicating orthogonal) 
of the estimated projection directions from CAP \citep{Zhao2021} (in their first four leading components) 
and those from the proposed Bayesian latent factor model, 
which shows positive association for each projection direction  
with the similarity at least $0.4$.  
We also report the CAP regression coefficients (with 95\% bootstrap confidence intervals) for each estimated projected outcome component. 

According to the meta analysis in \cite{Smith2009}, 
the identified Parcel Set contrasting 
the Short vs. Conventional sleeper 
in Figure~\ref{fig.VR}    
mainly correspond to visual areas (network nodes N1, N3, N4, N8), auditory areas (N12, N13) and sensorymotor (N11). 
\cite{Curtis2016} found that 
self-reported sleep duration primarily co-varied with FC in auditory, visual, and sensorimotor cortices. 
Specifically, shorter sleep durations were associated with increased FC between between auditory, visual, and sensorimotor cortices (these regions roughly correspond to the network nodes N1, N3, N4, N8, N12, N13, and N11), and decreased FC between these regions and the cerebellum (N9). 
These positive and negative associations found in \cite{Curtis2016} are consistent with the results in the contrast maps presented in Figure~\ref{fig.offdiag} 
which contrast Short vs. Conventional sleepers.

 \section{Discussion}

 Extending the frequentist approach developed in \cite{Zhao2021} under a 
 probabilistic model (\ref{lfm}), 
 coupled with a  
 geometric formulation of the 
 dimension-reduced covariance objects $\bm{\Psi}_i$ in (\ref{cov.model}), 
    the proposed Bayesian method 
    provides a framework to 
    conduct
     inference on all relevant  parameters simultaneously, that produces more interpretable results 
regarding how the covariates' effects are expressed in the ROIs. 
Furthermore, the outcome dimension reduction approach avoids the need to work with subject-specific full $p$-by-$p$ sample covariance matrices, which can suffer from estimation instability when the number of time points (volumes) is not large (which is typically the case for fMRI signals). 
Generally, 
 the CAP formulation of
  \cite{Zhao2021} 
allows for a more targeted and efficient analysis by identifying the specific components of the outcome data 
 relevant to the association between covariates and FC.  

Although  the computational burden and complexity associated with working with the full $p$-by-$p$ sample covariance matrix can be significantly alleviated by reducing the dimensionality of the outcome data, 
 the method is generally  
 not suitable to be run in very high-dimensional outcome data, such as voxel-level data, and 
is better suited for intermediate spaces, such as those produced by ICA or an anatomical parcellation. 
 Overfitting might occur due to the large number of parameters in the estimation of the outcome projection matrix $\bm{\Gamma}$. 
Future work will apply prior distributions on the dimension reducing matrix $\bm{\Gamma}$ 
as well as on the covariate effect parameters $\bm{B}$ 
that promote sparsity, 
for improved estimation and interpretation 
in higher dimensional spaces. 
 
 As in \cite{Zhao2021,Zhao2021b, Zhao2022}, 
 the assumption that we make in conducting the inference 
  is partially common eigenvectors of the covariance structure \citep{Wang2021}, 
in which 
the covariance is decomposed into shared and unique components, 
 where
 the shared components 
  captures the information related to the covariates.  
Future endeavors will explore strategies to mitigate concerns related to model misspecification by addressing heterogeneity in these shared components across subjects.  
We have conducted preliminary thinning of the observed multivariate time-series to achieve an effective sample size, involving subsampling to eliminate temporal dependencies.  
 Subsequent investigations will refine this approach to delve into individual differences in dynamic FC  \citep[e.g.,][]{Zhang2020, 
 Bahrami2022},  incorporating dimension reduction models that account for both between-subject heterogeneity in spatial patterns and within-subject temporal correlation through state-space modeling of latent factors. 
  This will facilitate a deeper exploration of associations between covariates and FC.
  
   A main challenge in modeling covariance matrices is the positive definiteness constraint. 
Unlike a mean vector where a link function can act element-wise, the positive-definiteness on a covariance matrix is a constraint on all its entry \citep{Pourahmadi2011}.  
  One approach is to transform the problem into an unconstrained estimation problem through a transformation  such as Cholesky decomposition, although this requires natural ordering information.  
Alternative way is to consider 
a more fundamental geometric formulation, 
that views individual covariances  
 as elements on a (nonlinear) manifold.   
A more global transformation (compared to an entry-wise transformation) such as matrix log-transformation then maps individual covariances to a tangent space, allowing for unconstrained operations.  
However, a global log-transformation poses interpretability challenges, 
as it generally alters the covariate’s impact directions with respect to the measured ROIs.   
Our geometry-based CAP approach focuses on identifying relevant eigenvectors, while simultaneously 
estimating eigenvalues-by-covariates associations through a linear model in a tangent space. 
By assuming and identifying relevant eigenvectors $\bm{\Gamma}$ that align with the covariates’ impact directions, the global log transformation maintains their orientation regarding the covariates' effects, thus the estimated pairwise covariance contrasts preserve their interpretability as covariate-induced pairwise connectivity differences.  

Yet another important challenge 
is the high dimensionality, 
as the number of covariance elements increase quadratically in the response variable's dimension. 
Generally, CAP regression of \cite{Zhao2021}, and its extension developed here,  
is useful if there is no need to model the generation of the entire observations,  
and one is only interested in isolating the data into a potentially low-dimensional representation in which they exhibit certain desired characteristics such as maximizing the model likelihood associated with $\bm{x}_i$. 
Such supervised dimension reductions can generally mitigate the curse of dimensionality in covariance modeling.





\section*{Acknowledgements}

The author is grateful to Drs. Xiaomeng Ju and Thaddeus Tarpey for helpful discussions and to the three reviewers and the editors for their constructive reviews of this manuscript. 
This work was supported by National Institutes of Health (NIH) grant 5 R01 MH099003.
Data were provided by the Human Connectome Project, WU-Minn Consortium (Principal Investigators: David Van Essen and Kamil Ugurbil; 1U54MH091657) funded by the 16 NIH Institutes and Centers that support the NIH Blueprint for Neuroscience Research; and by the McDonnell Center for Systems Neuroscience at Washington University.

\section*{Supplementary Materials}
  The code used in this paper is accessible at the following GitHub repository: \\ \url{https://github.com/syhyunpark/bcap}.

\bibliographystyle{biorefs.bst}
\bibliography{refs}  

\end{document}